%% file: main.tex
\pdfoutput=1
\documentclass[a4paper,USenglish,cleveref,thm-restate]{lipics-v2021}
\usepackage[T1]{fontenc}
\usepackage{amsmath,amsfonts,amsthm}
\usepackage[]{todonotes}
\usepackage[ruled,linesnumbered,noend]{algorithm2e}
\usepackage{comment}
\usepackage{tikzit}

\usepackage[appendix=append]{apxproof} 
\usepackage[append]{optional}

\newtheoremrep{theorem}{Theorem}[section]
\newtheoremrep{claim}[theorem]{Claim}
\newtheoremrep{lemma}[theorem]{Lemma}
\newtheoremrep{corollary}[theorem]{Corollary}
\newtheoremrep{definition}[theorem]{Definition}
\newtheoremrep{remark}[theorem]{Remark}

\newcommand{\M}{\mathcal{M}}
\newcommand{\E}{\mathcal{E}}

\newcommand{\START}{\operatorname{START}}

\newcommand{\FIRE}{\operatorname{FIRE}}
\newcommand{\WAKE}{\operatorname{WAKE}}

\title{Network Abstractions for Characterizing Communication Requirements in Asynchronous Distributed Systems}

\titlerunning{Network Abstractions for Asynchronous Distributed Systems}



\nolinenumbers 

\hideLIPIcs  

\EventEditors{John Q. Open and Joan R. Access}
\EventNoEds{2}
\EventLongTitle{42nd Conference on Very Important Topics (CVIT 2016)}
\EventShortTitle{CVIT 2016}
\EventAcronym{CVIT}
\EventYear{2016}
\EventDate{December 24--27, 2016}
\EventLocation{Little Whinging, United Kingdom}
\EventLogo{}
\SeriesVolume{42}
\ArticleNo{23}

\author{Hugo Rincon Galeana}{TU Wien, Austria}{hugorincongaleana@gmail.com}{http://orcid.org/0000-0002-8152-1275}{}
\author{Ulrich Schmid}{TU Wien, Austria}{s@ecs.tuwien.ac.at}{http://orcid.org/0000-0001-9831-8583}{}

\authorrunning{H. Rincon Galeana, U. Schmid}

 		\funding{Hugo Rincon Galeana has been supported by the Doctoral College on Resilient Embedded Systems
	at TU Wien and the FWF project DMAC (P32431).
 	}

\begin{document}
  
\input{drawings/complexes.tikzstyles}

\maketitle

\begin{abstract}
Whereas distributed computing research has been very successful in exploring the solvability/impossibility border of distributed
computing problems like consensus in representative classes of computing models with respect to model parameters like failure bounds, this is not the case for characterizing necessary and sufficient communication 
requirements. In this paper, we introduce network abstractions as a novel approach for modeling communication requirements in asynchronous distributed systems. A network abstraction of a run is a sequence of directed graphs on 
the set of processes, where the $i$-th graph specifies some ``potential'' message chains that can be guaranteed to arise in the $i$-th portion of the run. Formally, they are defined via associating message sending times with 
the end-to-end delays that would arise if the message was indeed sent by the sender's protocol. Network abstractions also allow to reason about future causal cones that might arise in a run, hence also facilitate reasoning about liveness properties, and are inherently compatible with temporal epistemic reasoning frameworks.
We demonstrate the utility of our approach by providing necessary and sufficient network abstractions
for solving the canonical firing rebels with relay (FRR) problem, and variants thereof, in asynchronous message-passing systems with up to $f$ byzantine processes connected via point-to-point links. FRR is not only a basic primitive in clock synchronization and consensus algorithms, but also integrates several distributed computing problems, namely triggering events, agreement and even stabilizing agreement, in a single problem instance.
\end{abstract}

\input{1-introduction}
\input{2-prelim}
\input{3-results}

\input{conclusions}

\bibliographystyle{plain}
\bibliography{localbib}
\newpage

\appendix
\input{4-logic}
\end{document}

%% file: drawings/complexes.tikzstyles

\tikzstyle{VB}=[fill={rgb,255: red,236; green,255; blue,255}, draw=black, shape=circle, thick, minimum size=6mm, inner sep=0pt, font={\footnotesize}, shading=ball, ball color={{rgb,255: red,236; green,255; blue,255}}]
\tikzstyle{VG}=[fill=green, draw=black, shape=circle, thick, minimum size=6mm, inner sep=0pt, font={\footnotesize}, ball color=green]
\tikzstyle{VR}=[fill=red, draw=black, shape=circle, thick, minimum size=6mm, inner sep=0pt, font={\footnotesize}, ball color=red]
\tikzstyle{VY}=[fill=yellow, draw=black, shape=circle, thick, minimum size=6mm, inner sep=0pt, font={\footnotesize}, ball color=yellow]
\tikzstyle{VBsmall}=[fill={rgb,255: red,236; green,255; blue,255}, draw=black, shape=circle, thick, minimum size=2mm, inner sep=0pt, font={\footnotesize}, shading=ball, ball color={{rgb,255: red,236; green,255; blue,255}}]
\tikzstyle{VGsmall}=[fill=green, draw=black, shape=circle, thick, minimum size=2mm, inner sep=0pt, font={\footnotesize}, ball color=green]
\tikzstyle{VRsmall}=[fill=red, draw=black, shape=circle, thick, minimum size=2mm, inner sep=0pt, font={\footnotesize}, ball color=red]
\tikzstyle{VYsmall}=[fill=yellow, draw=black, shape=circle, thick, minimum size=2mm, inner sep=0pt, font={\footnotesize}, ball color=yellow]
\tikzstyle{VBfaded}=[fill={rgb,255: red,236; green,255; blue,255}, draw=black, shape=circle, thick, minimum size=2mm, inner sep=0pt, font={\footnotesize}, opacity=0.4]
\tikzstyle{VGfaded}=[fill=green, draw=black, shape=circle, thick, minimum size=2mm, inner sep=0pt, font={\footnotesize}, opacity=0.4]
\tikzstyle{VRfaded}=[fill=red, draw=black, shape=circle, thick, minimum size=2mm, inner sep=0pt, font={\footnotesize}, opacity=0.4]
\tikzstyle{VYfaded}=[fill=yellow, draw=black, shape=circle, thick, minimum size=2mm, inner sep=0pt, font={\footnotesize}, opacity=0.4]
\tikzstyle{VRBGsquare}=[top color={rgb,255: red,236; green,255; blue,255}, bottom color=green, middle color=red, draw=black, shape=rectangle, thick, minimum size=3mm, inner sep=0pt, font={\footnotesize}]
\tikzstyle{VBGsquare}=[top color={rgb,255: red,236; green,255; blue,255}, bottom color=green, draw=black, shape=rectangle, thick, minimum size=3mm, inner sep=0pt, font={\footnotesize}]
\tikzstyle{VGRsquare}=[top color=green, bottom color=red, draw=black, shape=rectangle, thick, minimum size=3mm, inner sep=0pt, font={\footnotesize}]
\tikzstyle{VRBsquare}=[top color=red, bottom color={rgb,255: red,236; green,255; blue,255}, draw=black, shape=rectangle, thick, minimum size=3mm, inner sep=0pt, font={\footnotesize}]
\tikzstyle{VBYsquare}=[top color={rgb,255: red,236; green,255; blue,255}, bottom color=yellow, draw=black, shape=rectangle, thick, minimum size=3mm, inner sep=0pt, font={\footnotesize}]
\tikzstyle{VGYsquare}=[top color=green, bottom color=yellow, draw=black, shape=rectangle, thick, minimum size=3mm, inner sep=0pt, font={\footnotesize}]
\tikzstyle{VRYsquare}=[top color=red, bottom color=yellow, draw=black, shape=rectangle, thick, minimum size=3mm, inner sep=0pt, font={\footnotesize}]
\tikzstyle{VBsquare}=[fill={rgb,255: red,236; green,255; blue,255}, draw=black, shape=rectangle, thick, minimum size=3mm, inner sep=0pt, font={\footnotesize}]
\tikzstyle{VGsquare}=[fill=green, draw=black, shape=rectangle, thick, minimum size=3mm, inner sep=0pt, font={\footnotesize}]
\tikzstyle{VRsquare}=[fill=red, draw=black, shape=rectangle, thick, minimum size=3mm, inner sep=0pt, font={\footnotesize}]
\tikzstyle{VYsquare}=[fill=yellow, draw=black, shape=rectangle, thick, minimum size=3mm, inner sep=0pt, font={\footnotesize}]
\tikzstyle{VBGfsquare}=[top color={rgb,255: red,236; green,255; blue,255}, bottom color=green, draw=black, shape=rectangle, ultra thick, minimum size=3mm, inner sep=0pt, font={\footnotesize}]
\tikzstyle{VGRfsquare}=[top color=green, bottom color=red, draw=black, shape=rectangle, ultra thick, minimum size=3mm, inner sep=0pt, font={\footnotesize}]
\tikzstyle{VRBfsquare}=[top color=red, bottom color={rgb,255: red,236; green,255; blue,255}, draw=black, shape=rectangle, ultra thick, minimum size=3mm, inner sep=0pt, font={\footnotesize}]
\tikzstyle{VBYfsquare}=[top color={rgb,255: red,236; green,255; blue,255}, bottom color=yellow, draw=black, shape=rectangle, ultra thick, minimum size=3mm, inner sep=0pt, font={\footnotesize}]
\tikzstyle{VGYfsquare}=[top color=green, bottom color=yellow, draw=black, shape=rectangle, ultra thick, minimum size=3mm, inner sep=0pt, font={\footnotesize}]
\tikzstyle{VRYfsquare}=[top color=red, bottom color=yellow, draw=black, shape=rectangle, ultra thick, minimum size=3mm, inner sep=0pt, font={\footnotesize}]
\tikzstyle{VBfsquare}=[fill={rgb,255: red,236; green,255; blue,255}, draw=black, shape=rectangle, ultra thick, minimum size=3mm, inner sep=0pt, font={\footnotesize}]
\tikzstyle{VGfsquare}=[fill=green, draw=black, shape=rectangle, ultra thick, minimum size=3mm, inner sep=0pt, font={\footnotesize}]
\tikzstyle{VRfsquare}=[fill=red, draw=black, shape=rectangle, ultra thick, minimum size=3mm, inner sep=0pt, font={\footnotesize}]
\tikzstyle{VYfsquare}=[fill=yellow, draw=black, shape=rectangle, ultra thick, minimum size=3mm, inner sep=0pt, font={\footnotesize}]
\tikzstyle{VRBGssquare}=[top color={rgb,255: red,236; green,255; blue,255}, bottom color=green, middle color=red, draw=black, shape=rectangle, thick, minimum size=2mm, inner sep=0pt, font={\footnotesize}]
\tikzstyle{VBGssquare}=[top color={rgb,255: red,236; green,255; blue,255}, bottom color=green, draw=black, shape=rectangle, thick, minimum size=2mm, inner sep=0pt, font={\footnotesize}]
\tikzstyle{VGRssquare}=[top color=green, bottom color=red, draw=black, shape=rectangle, thick, minimum size=2mm, inner sep=0pt, font={\footnotesize}]
\tikzstyle{VRBssquare}=[top color=red, bottom color={rgb,255: red,236; green,255; blue,255}, draw=black, shape=rectangle, thick, minimum size=2mm, inner sep=0pt, font={\footnotesize}]
\tikzstyle{VBYssquare}=[top color={rgb,255: red,236; green,255; blue,255}, bottom color=yellow, draw=black, shape=rectangle, thick, minimum size=2mm, inner sep=0pt, font={\footnotesize}]
\tikzstyle{VGYssquare}=[top color=green, bottom color=yellow, draw=black, shape=rectangle, thick, minimum size=2mm, inner sep=0pt, font={\footnotesize}]
\tikzstyle{VRYssquare}=[top color=red, bottom color=yellow, draw=black, shape=rectangle, thick, minimum size=2mm, inner sep=0pt, font={\footnotesize}]
\tikzstyle{VBssquare}=[fill={rgb,255: red,236; green,255; blue,255}, draw=black, shape=rectangle, thick, minimum size=2mm, inner sep=0pt, font={\footnotesize}]
\tikzstyle{VGssquare}=[fill=green, draw=black, shape=rectangle, thick, minimum size=2mm, inner sep=0pt, font={\footnotesize}]
\tikzstyle{VRssquare}=[fill=red, draw=black, shape=rectangle, thick, minimum size=2mm, inner sep=0pt, font={\footnotesize}]
\tikzstyle{VYssquare}=[fill=yellow, draw=black, shape=rectangle, thick, minimum size=2mm, inner sep=0pt, font={\footnotesize}]

\tikzstyle{Eout}=[->, >=latex]
\tikzstyle{Ein}=[<-, >=latex]
\tikzstyle{Enone}=[-]
\tikzstyle{Einout}=[<->, >=latex]
\tikzstyle{Eoutthick}=[->, >=latex, very thick]
\tikzstyle{Einthick}=[<-, >=latex, very thick]
\tikzstyle{Enonethick}=[-, very thick]
\tikzstyle{Einoutthick}=[<->, >=latex, very thick]
\tikzstyle{Enonedotted}=[-, dash dot]
\tikzstyle{Eoutd}=[->, >=latex, dotted, thin]
\tikzstyle{Eind}=[<-, >=latex, dotted, thin]
\tikzstyle{Enoned}=[-, dotted, thin]
\tikzstyle{Einoutd}=[<->, >=latex, dotted, thin]
\tikzstyle{FaceBlue}=[-, fill={rgb,255: red,43; green,188; blue,255}, draw=none, opacity=0.4]
\tikzstyle{FaceYellow}=[-, fill={rgb,255: red,255; green,255; blue,119}, draw=none, opacity=0.4]
\tikzstyle{ArrowBlack}=[-, fill=black]
\tikzstyle{FacePur}=[-, fill={rgb,255: red,124; green,37; blue,255}, draw=none, opacity=0.4]
\tikzstyle{EdgeBlue}=[-, draw=blue, very thick]
\tikzstyle{EdgeRed}=[-, draw=red, very thick]

%% file: 1-introduction.tex
\section{Introduction} \label{sec:intro}

A substantial part of the existing distributed computing research is devoted to the question of 
when some distributed computing problem, say, consensus \cite{LSP82}, is solvable in a given model 
of computation. Impossibility results, like the celebrated FLP consensus impossibility 
in asynchronous systems where just one process may crash \cite{FLP85}, tell when this is not possible. 
Otherwise, solution algorithms, along with their correctness proofs, are provided.

Proving impossibility results is easier in weaker models of computation, where the adversary has more power. In addition, impossibility results a fortiori carry over to weaker computing models. For example, the FLP impossibility implies that consensus is also impossible if $f\geq 1$ processes may crash. Consequently, one usually tries to establish an impossibility result for a computing model that is as strong as possible. By contrast, designing and proving correct algorithms is easier in strong computing models, with the solutions remaining valid in stronger models only. Consequently, it is desirable to find algorithms for computing models that are as weak as possible. Sometimes, the respective attempts can be made matching, which allows one to precisely identify the possibility/impossibility border. For example, in synchronous systems with up to $f$ byzantine faulty processes, it is known that $n=3f+1$ processes are necessary and sufficient for solving consensus \cite{LSP82}.

Unfortunately, exploring this possibility/impossibility border with respect to communication requirements 
is much harder. On the one hand, there is a substantial body of distributed computing research devoted to information-theoretic 
communication complexity \cite{KN97}, which has been sparked by Yao's seminal work \cite{Yao79} on distributed function computation for two processes. A few classic examples, among many possible others, are symmetry breaking in 
chains and rings \cite{DMR08} and lower bounds for all-pair shortest paths \cite{HW12:PODC}. However,
knowing the bit complexity of some problem does not necessarily allow one to assess whether it is possible 
to solve it some specific computer network. Moreover, analyzing the communication complexity of complex distributed
computing problems, like consensus for more than 2 processes, is inherently difficult \cite{PS18:SIROCCO}. 

Consequently, most traditional research on complex distributed computing problems 
considered networks of processes that are fully connected 
by means of bidirectional links that are reliable \cite{FLP85} or fair-lossy \cite{BCT96}.
In not fully-connected but still static network topologies, solving consensus 
with up to $f$ byzantine processes requires a $2f+1$-connected undirected communication graph
\cite{FLM86}, and some more involved connectivity constraints for directed communication graphs \cite{TV15:PODC}. 
Time-varying communication graphs \cite{CFQS12:TVG} are considered in research on dynamic networks \cite{KO11:SIGACT}, 
where it is usually assumed that a message adversary \cite{AG13} supplies the directed communication graph 
for every particular round. 
Both impossibility results and solution algorithms have been provided for a wide variety of different 
message adversaries, ranging from oblivious ones \cite{COULOUMA201580,winkler_et_al:LIPIcs.ITCS.2023.100} to 
ones that eventually ensure a short period of stability only \cite{WSS19:DC}. Albeit most of this 
research focuses on consensus, other distributed computing problems like $k$-set agreement \cite{GP16:OPODIS,GWSR19:SSS} and stabilizing consensus \cite{CM19:DC,SS21:SSS} have been studied in such models as well.

Unfortunately, most solution algorithms depend heavily on the underlying system assumptions,
which are often incomparable, in particular, in the case of message adversaries. Determining 
necessary and sufficient communication requirements for solving a given problem, which
are reasonably independent of the underlying model, and thus facilitate a rating of the
overall communication efficiency of a particular solution, appears rather hopeless. 
Indeed, besides the already mentioned connectivity results \cite{FLM86,TV15:PODC} 
for distributed systems with static communication topologies, we are only aware of very few attempts on characterizing the 
problem solvability/impossibility border for dynamic networks under message adversaries in general, 
which all rely on very abstract properties: In the case of consensus, Nowak, Schmid and Winkler
\cite{10.1145/3293611.3331624} showed that the non-connectedness/connectedness 
of the topological space of the infinite admissible executions generated by a solution algorithm defines the border. Whereas this condition is far from being practical, it can also be 
expressed in terms of broadcastability/non-broadcastability of the connected components
of the space, which has been exploited in the combinatorial characterization of
limit-closed message adversaries in \cite{WSM19:OPODIS}. 
In \cite{SSW18:SIROCCO}, Schmid, Schwarz and Winkler introduced message adversary 
simulations, which define a message adversary hierarchy and the notion of a strongest 
message adversary for solving a given distributed computing problem. The discriminating power of the
latter approach is low, however,
in the sense that many incomparable message adversaries are usually strongest ones. 
In any case, such characterizations are way too abstract for distilling necessary and sufficient 
communication requirements out of those.

By contrast, exactly this is possible in the two-step approach advocated by Ben-Zvi and Moses
in \cite{BM14:JACM}. The authors considered the simple \emph{ordered response} (OR) problem 
in a distributed computing model with asynchronous but time-bounded end-to-end communication here. To solve OR, the
processes must execute some action $\FIRE$, in a fixed sequence, when some trigger event $\START$ 
has occurred at some process. First, epistemic analysis \cite{HM90} was used to provide the necessary 
and sufficient knowledge that must be acquired by the processes in every run to correctly solve OR. 
In a second step, the necessary and sufficient amount of communication needed for establishing this knowledge 
was determined: 
Since the underlying computing model facilitates communication-by-time \cite{Lam78}, i.e., replacing
the sending of a message representing some fact $\varphi$ by not sending any message and using a timeout 
at the receiver to learn about $\varphi$, a communication pattern
called a \emph{centipede}, which generalizes the usual message chains in purely asynchronous systems, 
must occur in every run.

\medskip

In this paper,\footnote{A brief announcement of a preliminary version of this paper has been accepted at SIROCCO'24.} we propose a novel approach for characterizing necessary and sufficient communication
requirements for asynchronous message-passing systems: \emph{network abstractions}. In a nutshell,
a network abstraction governing a run is a finite or infinite sequence $G_1,G_2,\dots$ of directed graphs on 
the set of processes, which defines \emph{some} of the \emph{potential} causal message chains that are guaranteed to occur in the run. 
The index $i$ in the graph sequence $G_1,G_2,\dots$ represents the progress of time, and $G_i$ specifies the potential message 
chains that must occur in the $i$-th portion of the run. Whereas these portions could be viewed as very loosely synchronized asynchronous rounds, we avoid using this term and define them implicitly only.

Potential causal message chains can be viewed as a \emph{possibility} for multi-hop communication offered by the adversary, which, however, only results in a corresponding message chain in the run if the protocols of the involved processes also attempt to send a message at the appropriate times. Alternatively, however, a hop in such a chain could also be formed by \emph{not} sending a message and relying on communication-by-time \cite{Lam78,BM14:JACM,GM20:JACM} in time-bounded asynchronous systems instead. Note that potential causal message chains were called \emph{syncausal} in \cite{BM14:JACM}. Technically, we accomplish this by associating message sending times with the end-to-end delays that would arise if the message was indeed sent by the sender's protocol.
As a consequence, network abstractions also allow to reason about future causal cones ~\cite{EPTCS297.19} 
that might arise in a run, hence also facilitate reasoning about liveness properties.

To showcase the utility of our approach, we identify necessary and sufficient network abstractions for a (weak) agreement 
problem called \emph{firing rebels with relay} (FRR), and its simpler variant \emph{firing rebels} (FR), in purely asynchronous systems with up to $f$ byzantine processes connected by point-to-point links,
which has been introduced and epistemically analyzed by Fruzsa, Kuznets and Schmid in ~\cite{FireTark}. FRR requires all correct agents to execute an action $\FIRE$ in an all-or-nothing fashion when sufficiently many agents learned about the occurrence of sufficiently many $\START$ events at correct processes. Informally, its specification comprises the following three properties:
\begin{enumerate}
    \item[(C)] \emph{Correctness}: If sufficiently many processes observed $\START$, then all correct processes must execute $\FIRE$ eventually.
    \item[(U)] \emph{Unforgeability}: If a correct process executes $\FIRE$, then $\START$ was observed by a correct process.
    \item[(R)] \emph{Relay}: If a correct process executes $\FIRE$, then eventually all correct processes execute $\FIRE$.
\end{enumerate} 
For FR, only (C) and (U) need to hold.

The FRR problem is particularly suitable for our purpose, since it is a canonical problem that combines
several interesting aspects of distributed computing problems in a single instance: 
It is not only closely related to the firing 
squad problem~\cite{10.1145/22145.22182,CHARRONBOST2019100}, which is a weak terminating agreement problem, 
but is also the core of an information-less version of the pivotal \emph{consistent broadcasting} (CB) primitive
introduced by Srikanth and Toueg in \cite{ST87}. CB has been used in fault-tolerant clock synchronization
\cite{ST87,WS09:DC,RS11:TCS}, in byzantine synchronous consensus \cite{ST87,DLS88}, and (in a slightly 
extended form) in the simulation of crash-prone protocols in byzantine settings proposed in \cite{10.1145/2591796.2591853}. 
FRR itself is relevant in practice for (almost) simultaneously triggering an event at different processes, and is interesting theoretically also because of its relation to stabilizing consensus \cite{CM19:DC}, namely, in the case where no correct process ever executes $\FIRE$.


It seems appropriate here to compare our network abstractions to some classic modeling 
approaches that appear to be somehow related.

First of all, network abstractions are a generalization of the time-bounded communication
assumption used in \cite{BM14:JACM}. Moreover, analogous to the ordered response problem
studied there, we also established the necessary knowledge\footnote{It consists of some part related to common knowledge \cite{HM90} (which is inevitable for satisfying the agreement property) and the knowledge that $\START$ has occurred on some correct process. Note that the presence of byzantine agents does not allow one to use the standard
knowledge modality here, but rather requires epistemic modalities such as \emph{hope} and \emph{belief}, see \cref{sec:logic} for an introduction. Sufficient conditions for acquiring the latter have been established in~\cite{EPTCS379.37}.} that must be acquired by every correct process for solving FR and FRR \cite{FireTark}. We stress, however, that epistemic results do not reveal anything about necessary and sufficient communication requirements.

First of all, one might argue that network abstractions somehow resemble message adversaries~\cite{AG13}, 
in the sense that both approaches represent a time-varying network topology as directed graph sequences. 
A fundamental difference lies in what some directed graph $G_i$ represents, however. In the synchronous message adversary setting, it represents \emph{all} the successful point-to-point communication happening in round $i$. In our context, $G_i$ specifies \emph{some} real or potential message chains that might occur in the $i$-th portion of a run, which is defined implicitly via path concatenations of the earlier portions (see \cref{def:finnetabs}). In stark contrast to the message adversary setting, our portions are hence only very loosely synchronized and allow us to concentrate on the essential part of the 
communication needed for solving a problem. 

Time-varying graphs (TVG), as introduced in \cite{CFQS12:TVG}, have been developed with the incentive to describe all conceivable variants of distributed systems with time-varying communication topologies.
The basic definition of a TVG is hence very expressive and provides, in particular, a time-dependent latency 
function for edges that is equivalent to our delay map $\sigma$. The framework also provides the notion of a journey, which 
is the analog of a path-closed causal chain (see \cref{def:pathclosedsch}). These and other concepts have been used to characterize TVGs into a partial order of 10 different classes, which differ in their communication guarantees. Rather than identifying fine-grained finite or infinite network abstractions on a per run basis as we do, which is inevitable for studying the necessary and sufficient communication requirements for solving a certain problem, however, \cite{CFQS12:TVG} focuses on generic network properties like connectivity over time, recurrent connectivity or time-bounded recurrence of edges.

\subparagraph*{Summary of our contributions:} 
\begin{itemize}
\item A novel framework that enables us to represent some potential communication in an asynchronous message-passing system as sequences of directed graphs called network abstractions.
\item A characterization of the necessary and sufficient network abstractions for solving FR and FRR in purely asynchronous systems. In the appendix, we also sketch a temporal epistemic logic formulation of necessary and sufficient a-priori knowledge for solving FRR.
\item Novel asynchronous protocols for FR and FRR that, unlike the existing solutions, also work correctly 
under the necessary and sufficient network abstractions identified above.
\end{itemize}
Overall, our findings demonstrate that our novel approach indeed allows one to identify necessary and sufficient \emph{time-evolving} network connectivity conditions for solving a given distributed computing problem in a byzantine asynchronous system, and to develop optimal solution algorithms. If a given network does not provide at least this network connectivity, there is no chance to solve the problem. We hence claim that our approach is of both theoretical and practical interest. Moreover, whereas we cannot judge the applicability to other problems yet, we are not aware of any alternative approach that could be used to obtain similar results.

\subparagraph*{Paper organization:} 

 In the following \cref{sec:prelim}, we provide our system model, the general assumptions and the basic concepts used throughout our paper. In \cref{sec:results}, for both FR and FRR, we provide (1) solution algorithms and (2) a precise characterization of the communication abstractions that make these problems solvable. Our findings are complemented by a temporal epistemic logic formulation of our characterizations found in~\cref{sec:logic}. A short summary of our accomplishments and some ideas for future work in \cref{sec:conclusion} round off our paper.

%% file: 2-prelim.tex
\section{System Model} \label{sec:prelim}

We consider a finite set of processes $\Pi = \{ p_1, \ldots, p_n\}$ with unique ids, which execute a deterministic or non-deterministic protocol\footnote{Note that all the protocols provided in this paper will actually be deterministic: For exploring solvability/impossibility issues, we will employ flooding full-information protocols, where the processes record and broadcast their full history whenever they make a step. Moreover, some more practical protocols for FR and FRR with reasonable message size, memory and computing requirements will also be provided.} and communicate via message passing over point-to-point links. In every run, some subset $B \subseteq \Pi$ of up to $f \geq 1$ byzantine processes may deviate arbitrarily from the protocol. A byzantine process could hence omit sending messages, send messages with wrong information, and take arbitrary state transitions, but cannot impersonate other processes due to our point-to-point link assumption. Note that a byzantine process could also simulate the behavior of a correct process. Whereas the membership of a process in $B$ is assumed to be static and a priori unknown to the processes, the value of $f$ is usually commonly known. Since FR, which needs a majority of correct processes \cite{FireTark}, is the weakest problem considered in this paper, we will assume $n\geq 2f + 1$.

As usual, processes execute a protocol given as a state machine, the execution of which is controlled by an adversary. Whenever a process is scheduled to take a step, it applies its transition function to its current local state and some subset of the received but not yet processed messages to compute a new local state and a set of messages to be sent. We allow non-deterministic state transitions, albeit the \emph{range} of possible choices must depend on the current local state only; this ensures that a process with the same local state in different executions can take the same state transitions. The adversary also determines when a message is delivered to the receiver. The time interval between the step where a message is sent to the step where the receiver processes the delivered message is called \emph{end-to-end delay}, which we assume to be $> 0$ and finite in the case of purely asynchronous systems and bounded in the case of time-bounded asychronous systems. Processes may also experience distinguished \emph{external events} $E\in\mathbf{E}$ like $\START$, scheduled by the adversary, which occur at most once at every process and are processed in the subsequent state transition.

We do not generally require any synchrony properties other than the existence of a linear global time and Zeno-freedom: Every event in the system (a step of a process, a message delivery, or an external event) happens at some objective global time, taken from a linearly ordered time set $\mathbb{T}$ that is usually either $\mathbb{N}$ or $\mathbb{R}$, and at most finitely many events may occur in a finite time interval. Note that processes do not necessarily have access to global time or even an internal notion of time, other than their local state transition histories, albeit such means could be added e.g. for implementing communication-by-time in time-bounded asynchronous systems \cite{Lam78,BM14:JACM,GM20:JACM}. 

In general, we consider systems with asynchronous starts, as it is usually assumed for asynchronous systems, and sometimes also for partially synchronous and even synchronous systems \cite{WS07:DC,CM19:DC}. All processes are assumed to be in their initial state at time 0 already, but effectively start their execution, i.e., become active, at different times and are not a priori aware of the already active processes. Messages that are sent to a process that is not active at the time of the message delivery are lost without a trace. For convenience, we will assume that a process becomes active when the external event $\WAKE$ occurs.

System-wide, this leads to executions (also called runs) starting at time 0 that can be represented by the generated sequence of configurations, which are the vectors of the local states of all the processes and the set of still unprocessed delivered messages, taken at the global time when some process takes a step.

\subsection{Message and event schedules}
 
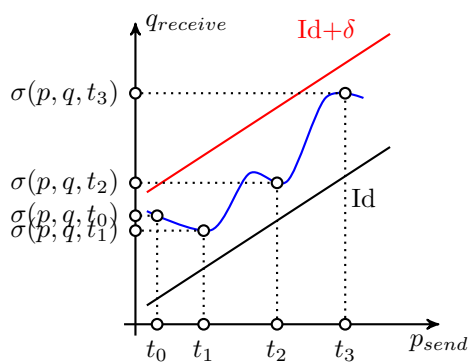
\begin{figure}[h]
	\centering
	\input{drawings/plot.tikz}
	\caption{A message delay map $\sigma$ (blue) that satisfies a bounded delay of $\delta$. $t_0, t_1, t_2, t_3$ are sending times from $p$ to $q$.}
    \label{fig:delmap}
\end{figure}

Contrasting traditional modeling, we specify the end-to-end delays of messages (potentially) sent in a run explicit by means of \emph{message delay maps}. Intuitively, a message delay map $\sigma$ tells at what time $\sigma(p,q,t)$ a process $q$ does or \emph{would} process a message received from process $p$ \emph{if} such a message was sent to it at time $t$. Note that $\sigma$ is similar to the edge latency function of a time-varying graph as introduced in \cite{CFQS12:TVG}. A simple example is $\sigma(p,q,t)=t+\delta_{pq}$, as used in the time-bounded asynchronous model employed in \cite{BM14:JACM}, where $\delta_{pq}$ is an a priori known bound on the end-to-end delay of the messages from $p$ to $q$. General message delay maps like the one illustrated in \cref{fig:delmap} extend this to message delay bounds that can vary with time, e.g., caused by network bandwidth fluctuations or outages, as well as by variations caused by the protocol execution itself, due to faulty processes, message sizes, receiver queueing effects, network congestion etc.

It is important to note that $\sigma(p,q,t)$ must hold also if the sender $p$ is byzantine
faulty: if $p$ sends a (possibly erroneous) message to $q$, its end-to-end delay in the run must be $\sigma(p,q,t)$. Only a byzantine faulty receiver process $q$ is allowed to violate $\sigma(p,q,t)$, as it may pretend to have received and processed 
the corresponding message at an arbitrary time.

 \begin{definition}[Message delay map and schedule] \label{def:schedule}
    We say that $\sigma: \Pi \times \Pi \times \mathbb{T} \rightarrow \mathbb{T}\cup \{\infty\}$ is a \emph{message delay map} for some run $r$ (resp.\ some part $[t_0,t_1]$ of $r$), if for any $p,q \in \Pi$ and $t \in \mathbb{T}$ (resp. $t \in [t_0,t_1]$), it holds that $\sigma(p,q,t) > t $. We write $\sigma(p,q,t) = \infty$ if a message sent from $p$ to $q$ at time $t$ will never be delivered. The \emph{message schedule} $\M \subseteq \Pi \times \Pi \times \mathbb{T}$ governing the run $r$ (resp.\ some part $[t_0,t_1]$ of $r$) is the set of all tuples $(p,q,t)$, $t \in \mathbb{T}$ (resp. $t \in [t_0,t_1]$), where process $p$ either does send a message to $q$ or \emph{could} send such a message when employing communication-by-time \cite{Lam78,BM14:JACM,GM20:JACM}. 
 \end{definition}

Whereas there are usually many message delay maps $\sigma$ for some run $r$, which all provide the particular end-to-end delay of every message actually or potentially sent in the run but could be different at other times, the message schedule $\M$ governing $r$ is unique. We will therefore also say that $\M$ \emph{corresponds} to $r$, and will usually use $\M_\sigma$ to denote the message schedule $\M$ governing $r$ with delay map $\sigma$. Note carefully that $\M_\sigma$ subsumes both \emph{process-time graphs} \cite{BM14:JACM,EPTCS297.19}, which describe actual message chains in a run, and the \emph{bounds relation} in \cite{BM14:JACM}, which specifies potential message receive times and is used for defining \emph{syncausal} message chains. Clearly, any message schedule governing a run can contain at most be countable, and any message schedule governing a finite part of a run can at most be finite, due to our Zeno-freedomness assumption. 

We now turn our attention to external events, which are traditionally considered instantaneous \cite{BM14:JACM}. For the sake of a compact notation, we will subsequently denote external events just as events. In our setting, they are generalized to have a starting time and an ending time. The starting time can be viewed as the \emph{occurrence time} of the external event at a process $p$, whereas the ending time represents the time when $p$ has processed it, i.e., when it has acknowledged the occurrence of the event via its state machine. We say that the latter causes $p$ to \emph{witness} the external event, and call the time difference between occurrence and witnessing at a process the \emph{event delay}. 

An advantage of this modeling is that an event at process $p$ can now be treated as a self-message $p \to p$, sent by $p$ at time $t_i$ and processed by $p$ at time $t_f$. 
 
\begin{definition}[Event delay map and schedule]\label{def:eventsched}
    Let $P\subseteq \Pi$ be a set of processes. We say that $\sigma_E: P \rightarrow \mathbb{T} \times \mathbb{T}$ is an \emph{event delay map} governing some run $r$ (resp.\ some part $[t_0,t_1]$ of $r$) for the external event $E \in \mathbf{E}$, if $\sigma_E(p) = (t_i,t_f)$ whenever $E$ occurs at process $p$ at time $t_i \in \mathbb{T}$ (resp.\ $t_i \in [t_0,t_1]$) and has been witnessed at time $t_f > t_i$. Note that, if $t_f=\infty$, then the event is never witnessed by $p$ and would hence been considered as not having happened. The \emph{event schedule} $\E_E \subseteq \Pi \times \Pi \times \mathbb{T}$ governing the run $r$ (resp.\ some part $[t_0,t_1]$ of $r$) for the event $E$ is the set of all tuples $(p,p,t)$, $t \in \mathbb{T}$ (resp. $t \in [t_0,t_1]$), where event $E$ occurs at process $p$ at time $t$. 
\end{definition}
Again, we denote by $\E_{\sigma_E}$ the event schedule $\E_E$ governing $r$ for event $E$ with event map $\sigma_E$.

Clearly, end-to-end message delays and event delays depend not only on the network transmission delays and the external event occurrence times, but also on the actual protocol used, the sender and receiver process scheduling, and the behavior of faulty processes. As usual, we will assume that all these aspects, except for the non-deterministic choice occurring in a non-deterministic protocol, are under the control of the adversary. Whereas we allow the adversary to be adaptive in the latter case, its choices for event schedules must be independent of earlier message delays and schedules. Moreover, it must be oblivious 
w.r.t.\ byzantine faulty processes that behave like correct ones:

\begin{definition}[Fault-oblivious adversary]\label{def:adversary}
We assume that process scheduling, message delivery, external event occurrences as well as the choice of the byzantine processes and their behavior in a run is determined by an adversary, which may be \emph{adaptive} in the sense that it may choose the message and event delays and schedules governing the suffix of a run after time $t\geq 0$ to depend on the prefix of the run up to time $t$. The choice of the event schedules in the suffix must be independent of the earlier message delays and message schedules, however.
Moreover, the adversary must be \emph{fault-oblivious}, in the sense that if $\sigma$ governs some run $r$ where process $p$ is correct, and $r'$ is a run that is identical to $r$ except that $p$ is byzantine faulty but behaves exactly as $p$ behaves in $r$, then $\sigma$ must also govern $r'$. 
\end{definition}

\subsection{Network and communication abstractions}

In this section, we will use our representation of runs based on the corresponding message delays and schedules for defining a higher-level communication abstraction. In a nutshell, our \emph{network abstractions} will focus on \emph{causal message chains} $p_1 \to p_2 \to \dots \to p_\ell$, where $p_i \to p_{i+1}$ either represents $p_i$ sending a message to $p_{i+1}$ or else $p_i$ deliberately not sending a message and using communication-by-time instead, and will abstract away the particular timing properties. 
We introduce our approach for message schedules first, and explain later how to incorporate event delays and schedules.

A core property of a message schedule $\M_{\sigma}$ is \emph{path-closedness} with respect to some directed graph $G$ on $\Pi$. Informally, it guarantees that $\M_{\sigma}$ contains some causal message chain for every directed path in $G$. We will call $G$ a static network abstraction for $\M_{\sigma}$ in this case, where the term static reflects the fact that this type of network abstractions is not intended to cover any time evolution.

\begin{definition}[Path-closedness] \label{def:pathclosedsch}
	Let $\M_{\sigma}$ be a message schedule with delay map $\sigma$, and $G$ a directed graph with $V(G) = \Pi$.
	We say that $\M_\sigma$ is \emph{path-closed with respect to $G$}, if, for any simple directed path $\mathcal{P} = (x_1,  \ldots ,x_k)$ in $G$ (i.e., without repeating vertices), there exists a time sequence $(t_1, \ldots, t_{k-1})$ such that, for any $0 \leq i \leq k-1$, the following holds: 
 \begin{enumerate}
	    \item $(x_i,x_{i+1},t_i) \in \M_\sigma$ 
        \item$\sigma(x_i,x_{i+1},t_i) < t_{i+1}$ 
	\end{enumerate} 
\end{definition}

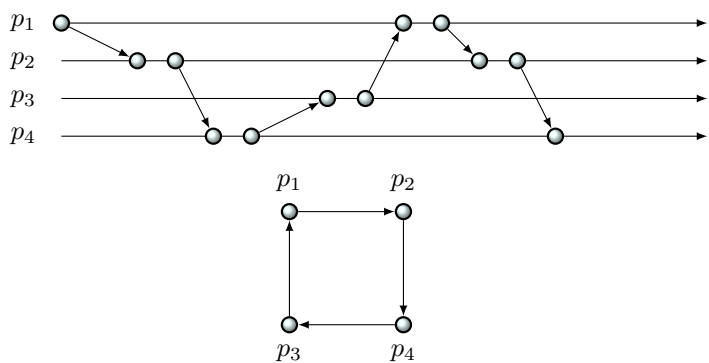
\begin{figure}[h]
	\centering
	\input{drawings/cycledyn.tikz}
	\caption{A process time graph and a corresponding static network abstraction.}
    \label{fig:staticna}
\end{figure}

\begin{definition}[Static network abstraction] \label{def:staticnet}
	Let $\M_{\sigma}$ be a message schedule with delay map $\sigma$ governing a run $r$. We say that a directed graph $G$ on $\Pi$ is a \emph{static network abstraction} for $r$ (as well as for $\M_{\sigma}$), or that $r$ adheres to (or has) a static network abstraction $G$, if $\M_{\sigma}$ is path-closed with respect to $G$.
\end{definition}
An example of a static network abstraction is shown in \cref{fig:staticna}.

The intuition behind a static network abstraction $G$ for a message schedule $\M_\sigma$, and hence for the run $r$ it governs, is that there is an \emph{opportunity} for relaying some information over the message chains described by $G$ in $r$. No information about \emph{when} this opportunity emerges is provided, however. Moreover, there is no guarantee that this opportunity emerges more than once. 
It should also be noted that, in general, static network abstractions for a run $r$ are neither unique nor complete, in the sense that they do not usually incorporate all message chains that exist in $r$. Furthermore, if both $G_1$ and $G_2$ are static network abstractions for $r$, $G_1 \cup G_2$ is not necessarily a static network abstraction for $r$. For example, $(p\to q) \in G_1$ and $(q\to r)\in G_2$ do not need to be consecutive in $\M_{\sigma}$, i.e., do not necessarily form the causal chain $(p\to q \to r)\in G_1 \cup G_2$. 

We will now extend static network abstractions in a way that makes them, despite their incompleteness and coarseness, appropriate for achieving our main goal, namely, identifying necessary and sufficient communication requirements for solving distributed computing problems in asynchronous systems. 

First, static network abstractions can be extended to finite graph sequences $\mathbf{G} = (G_1, \ldots, G_k)$, which allow to model the time evolution present in $\M_\sigma$ resp.\ in the run $r$ it governs. Intuitively, $r$ is chopped into $k$ consecutive portions here, where $G_i$ forms a static network abstraction for the $i$-th portion. This is formalized in the following \cref{def:finnetabs} by means of path concatenation of the simple paths in $(G_1, \ldots, G_k)$, which is defined as follows: If $\mathbf{P}_1$ and $\mathbf{P}_2$ denote the sets of
simple paths of $G_1$ and $G_2$, respectively, their path concatenation is defined as $\mathbf{P}_1 \oplus \mathbf{P}_2 = \{(x_1,x_2,\dots,x_m)| (x_1,\dots,x_k) \in \mathbf{P}_1, (x_{k},\dots,x_m) \in \mathbf{P}_2, 1 < k < m, \mbox{$x_i\neq x_j$ for $1\leq i,j\neq i \leq m$}\}$. 

\begin{figure}[h]
	\centering
	\input{drawings/cycle43.tikz}
	\caption{A process time graph of a run with static network abstractions $G_1$ and $G_2$. However, neither $G_1 \cdot G_2$ nor $G_2 \cdot G_1$ are finite network abstractions.}
 \label{fig:nonuniquestaticNA}
\end{figure}
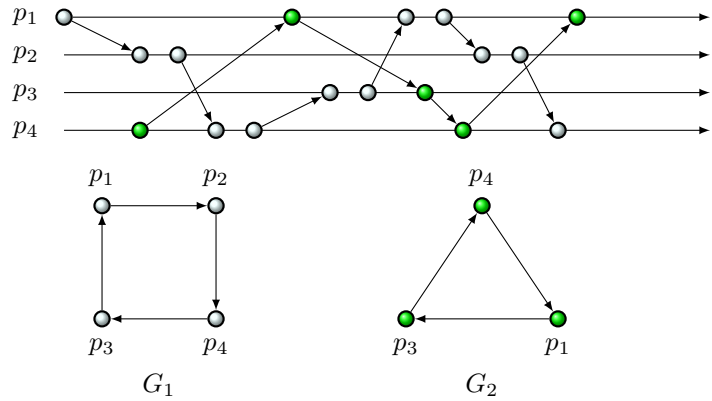

\begin{definition}[Finite network abstraction] \label{def:finnetabs}
	Let $G_1,\ldots, G_k$ be static network abstractions of the same message schedule $\M_{\sigma}$ governing a run $r$, and $\mathbf{P}_1, \ldots, \mathbf{P}_k$ their corresponding sets of simple paths. We say that $G_1 \cdot G_2 \cdots G_k$ is a \emph{finite network abstraction} for $r$ (as well as for $\M_{\sigma}$), or that $r$ adheres to (or has) a finite network abstraction $G_1 \cdot G_2 \cdots G_k$ if, for any simple path $P = (x_1, \ldots, x_m) \in \mathbf{P}_1 \oplus \ldots \oplus \mathbf{P}_k$, there is a time sequence $(t_1, \ldots, t_{m})$ such that for any $1 \leq i \leq m-1$:
    \begin{enumerate}
	    \item $(x_i,x_{i+1},t_i) \in \M_\sigma$
        \item$\sigma(x_i,x_{i+1},t_i) < t_{i+1}$
	\end{enumerate} 
 \end{definition}
An example of a finite network abstraction for a run is provided in \cref{fig:finiteNA}. By contrast, \cref{fig:nonuniquestaticNA} 
shows an example of two static network abstractions that do not compose into a finite network abstraction.

\begin{figure}[h]
	\centering
	\input{drawings/nocut.tikz}
	\caption{A process time graph of a run with a finite network abstraction $G_1 \cdot G_2$. Note that the sequence is determined by the causal chains, and that there is no time cut that splits the communication into $G_1$ and $G_2$.}
    \label{fig:finiteNA}
\end{figure}
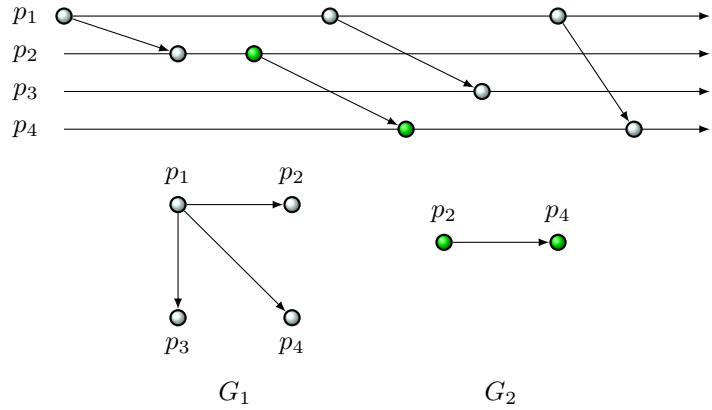

The following \cref{def:finnetabs} is a consequence of the fact that
causal paths occurring in the network abstraction governing
a run must occur in the run.

\begin{lemma}\label{rem:runprefix}
If a run $r$ has some finite network abstraction $G_1 \cdot G_2 \cdots G_k$, finite prefixes of $r$ need not adhere to it. 
\end{lemma}
\begin{proof}
The statement follows immediately from \cref{def:finnetabs} and \cref{def:schedule} of the message schedule for a part of a run, namely, for any prefix $[0,t_1]$ with $t_1 < t_{m}$ as defined in \cref{def:finnetabs}.
\end{proof}

Finite network abstractions can finally be extended to infinite sequences in a natural way, 
by considering all finite prefixes. The resulting \emph{dynamic network abstractions} cover infinite time evolution:

\begin{definition}[Dynamic network abstraction] \label{def:dynnetabs}
	Let $\mathcal{G} = G_1, \ldots, G_i, \ldots$ be an infinite graph sequence. We say that $\mathcal{G}$ is a \emph{dynamic network abstraction} for a run $r$, or that $r$ adheres to (or has) a dynamic network abstraction $\mathcal{G}$, if any finite prefix $P$ of $\mathcal{G}$ is a finite network abstraction for $r$.
\end{definition}

\begin{lemma}\label{rem:runsuffix}
If a run $r$ has a dynamic network abstraction $\mathcal{G} = G_1, \ldots, G_i, \ldots$, for every finite network abstraction $G_1 \cdot G_2 \cdots G_k$, there is a suffix of the run $r$ that does not adhere to it. 
\end{lemma}
\begin{proof}
The statement follows immediately from \cref{def:finnetabs} and \cref{def:schedule} of the message schedule for a part of a run, namely, for some suffix $[t_0,\infty)$ with $t_0 > t_{m}$ as defined in \cref{def:finnetabs}.
\end{proof}

\begin{remark}\label{rem:graphcompressor}
If finitely many graphs $\{G_1, \ldots, G_k\}$ all repeat infinitely often in a network abstraction $\mathcal{G}$ for a run $r$, then $G^\omega=G,G,\dots$ is also a dynamic network abstraction for $r$, where $G= G_1 \cup \ldots \cup G_k$ is the union graph consisting of all edges of $G_1, \ldots, G_k$. After all, it suffices to chop $\mathcal{G}$ into chunks that contain at least one instance of every $\{G_1, \ldots, G_k\}$. For every such chunk, $G$ is a static network abstraction.
\end{remark}

We will now generalize our network abstractions to also incorporate events, which will lead to \emph{communication abstractions}. Recall that we advocated already earlier to think of events as self-messages. Consequently, we can encode the occurrences and the witnessing of event $E$ as graphs consisting of loops only, with the convention that a loop counts as a simple path. This allows us to seamlessly interweave events and message chains.

\begin{definition}[Event graph] \label{def:eventgraph}
    Let $E$ be an event that is witnessed by the participating set of processes $S_E \subseteq \Pi$ in the event schedule $\E{\sigma_{E}}$ governing a run $r$. The corresponding \emph{event graph} $G_E = \langle V(G_E), E(G_E) \rangle$ for $E$ is defined by $V(G_E) = \Pi$, and $E(G_E) = \{ (p,p) \; \vert \; p \in S_E \}$.
\end{definition}

\begin{definition}[Finite communication abstraction]\label{def:commabs}
Let $\M_{\sigma}$ resp.\ $\E_{\sigma_{E}}$, $E\in\mathbf{E}$, be the message schedule resp.\ event schedules governing a run $r$.
Moreover, let $G_1,\ldots, G_k$ be a sequence consisting of either static network abstractions or event graphs for $r$, with $\mathbf{P}_1, \ldots, \mathbf{P}_k$ denoting their corresponding sets of simple paths (with loops considered as simple paths). We say that $G_1 \cdot \ldots \cdot G_k$ is a \emph{finite communication abstraction} for $r$, or that $r$ adheres to (or has) a 
finite communication abstraction $G_1 \cdot \ldots \cdot G_k$, if, for any simple path $P = (x_1, \ldots, x_m) \in \mathbf{P}_1 \oplus \ldots \oplus \mathbf{P}_k$, there is a finite time sequence $(t_1, \ldots, t_m)$ such that for $1 \leq i \leq m-1$:
    \begin{enumerate}
	    \item If $x_i \neq x_{i+1}$, then $(x_i,x_{i+1},t_i) \in \M_{\sigma}$
        \item If $x_i \neq x_{i+1}$, then $\sigma(x_i,x_{i+1},t_i) < t_{i+1}$ 
        \item If $x_{m-1} \neq x_m$, then $\sigma(x_{m-1},x_m, t_{m-1}) < \infty$
        \item If $x_i=x_{i+1}$, representing event $E$, then $\sigma_E(x_i) = (t_i, t')$ and $t' < t_{i+1}$
	\end{enumerate} 
 \end{definition}

\begin{definition}[Dynamic communication abstraction]\label{def:dyncomm}
    Let $\mathcal{G}$ be an infinite graph sequence consisting of either network abstractions or event graphs for a run $r$. We say that $\mathcal{G}$ is a \emph{dynamic communication abstraction} for $r$, or that $r$ adheres to (of has) a dynamic communication abstraction $\mathcal{G}$, if any finite prefix $P$ of $\mathcal{G}$ is a finite communication abstraction 
    for $r$.
\end{definition}

Recall that we assumed that the processes become active, i.e., start executing, when they have witnessed the external event $\WAKE$, which is assumed to eventually occur at every process. We can thus safely assume that every communication abstraction starts with the $\WAKE$ event graph. For conciseness, it will be omitted throughout the rest of our paper.

%% file: drawings/plot.tikz
\begin{tikzpicture}[
    thick,
    >=stealth',
    dot/.style = {
      draw,
      fill = white,
      circle,
      inner sep = 0pt,
      minimum size = 4pt
    }
  ]
  \coordinate (O) at (0,0);
  \draw[->] (-0.3,0) -- (8,0) coordinate[label = {below:$p_{send}$}] (xmax);
  \draw[->] (0,-0.3) -- (0,8) coordinate[label = {right:$q_{receive}$}] (ymax);
  \path[name path=x] (0.3,1.5) -- (6.7,5.7);
  \path[name path=y] plot[smooth] coordinates {(0.3,3) (2,2.5) (3,4) (4,3.8) (5,6) (6,6)};
  \path[name path=z] (0.3,2.7) -- (6.7,6.9);
  \scope[name intersections = {of = x and y, name = i}]
    \draw      (0.3,0.5) -- (6.7,4.7) node[pos=0.8, below right] {Id};
    \draw[red]      (0.3,3.5) -- (6.7,7.7) node[pos=0.9, above left] {Id$+\delta$};
    \draw[blue] plot[smooth] coordinates {(0.3,3) (2,2.5) (3,4) (4,3.8) (5,6) (6,6)};

    \draw[dotted] (i-3) -- (i-3 -| O) (i-3) -- (i-3 |-O) (i-1)--(i-1 -| O) (i-1)--(i-1 |- O) ;
    \draw(i-3) node[dot] (i-3) {}  (i-3 |- O) node[dot,
                              label = {below:$t_2$}] {};
    \draw (i-1) node[dot] {} (i-1 |- O) node[dot, label = {below:$t_1$}] {};
    \draw (i-1 -| O) node[dot, label = {left:$\sigma(p,q,t_1)$}] {};
    \draw (i-3 -| O) node[dot, label = {left:$\sigma(p,q,t_2)$}] {};
                              (i-12);
  \endscope
  \scope[name intersections = {of = y and z, name = i}]
    \draw[dotted] (i-3) -- (i-3 -| O) (i-3) -- (i-3 |-O) (i-1)--(i-1 -| O) (i-1)--(i-1 |- O) ;
    \draw(i-3) node[dot] (i-3) {}  (i-3 |- O) node[dot,
                              label = {below:$t_3$}] {};
    \draw (i-1) node[dot] {} (i-1 |- O) node[dot, label = {below:$t_0$}] {};
    \draw (i-1 -| O) node[dot, label = {left:$\sigma(p,q,t_0)$}] {};
    \draw (i-3 -| O) node[dot, label = {left:$\sigma(p,q,t_3)$}] {};
  \endscope
\end{tikzpicture}

%% file: drawings/cycledyn.tikz
\begin{tikzpicture}
	\begin{pgfonlayer}{nodelayer}
		\node [style=VBsmall] (0) at (-8, 6) {};
		\node [style=none] (1) at (-8, 5) {};
		\node [style=none] (2) at (-8, 4) {};
		\node [style=none] (3) at (-8, 3) {};
		\node [style=none] (4) at (9, 6) {};
		\node [style=none] (5) at (9, 5) {};
		\node [style=none] (6) at (9, 4) {};
		\node [style=none] (7) at (9, 3) {};
		\node [style=none] (8) at (-9, 6) {$p_1$};
		\node [style=none] (9) at (-9, 5) {$p_2$};
		\node [style=none] (10) at (-9, 4) {$p_3$};
		\node [style=none] (11) at (-9, 3) {$p_4$};
		\node [style=VBsmall] (12) at (-6, 5) {};
		\node [style=VBsmall] (13) at (-5, 5) {};
		\node [style=VBsmall] (14) at (-4, 3) {};
		\node [style=VBsmall] (15) at (-3, 3) {};
		\node [style=VBsmall] (16) at (-1, 4) {};
		\node [style=VBsmall] (17) at (0, 4) {};
		\node [style=VBsmall] (18) at (1, 6) {};
		\node [style=VBsmall] (19) at (2, 6) {};
		\node [style=VBsmall] (20) at (3, 5) {};
		\node [style=VBsmall] (24) at (4, 5) {};
		\node [style=VBsmall] (25) at (5, 3) {};
		\node [style=VBsmall] (26) at (-2, 1) {};
		\node [style=VBsmall] (27) at (1, 1) {};
		\node [style=VBsmall] (28) at (1, -2) {};
		\node [style=VBsmall] (29) at (-2, -2) {};
		\node [style=none] (30) at (-2, 1.75) {$p_1$};
		\node [style=none] (31) at (1, 1.75) {$p_2$};
		\node [style=none] (32) at (-2, -2.75) {$p_3$};
		\node [style=none] (33) at (1, -2.75) {$p_4$};
	\end{pgfonlayer}
	\begin{pgfonlayer}{edgelayer}
		\draw [style=Eout] (0) to (4.center);
		\draw [style=Eout] (1.center) to (5.center);
		\draw [style=Eout] (2.center) to (6.center);
		\draw [style=Eout] (3.center) to (7.center);
		\draw [style=Eout] (0) to (12);
		\draw [style=Eout] (13) to (14);
		\draw [style=Eout] (15) to (16);
		\draw [style=Eout] (17) to (18);
		\draw [style=Eout] (19) to (20);
		\draw [style=Eout] (24) to (25);
		\draw [style=Eout] (26) to (27);
		\draw [style=Eout] (27) to (28);
		\draw [style=Eout] (28) to (29);
		\draw [style=Eout] (29) to (26);
	\end{pgfonlayer}
\end{tikzpicture}

%% file: drawings/nocut.tikz
\begin{tikzpicture}
	\begin{pgfonlayer}{nodelayer}
		\node [style=VBsmall] (0) at (-8, 6) {};
		\node [style=none] (1) at (-8, 5) {};
		\node [style=none] (2) at (-8, 4) {};
		\node [style=none] (3) at (-8, 3) {};
		\node [style=none] (4) at (9, 6) {};
		\node [style=none] (5) at (9, 5) {};
		\node [style=none] (6) at (9, 4) {};
		\node [style=none] (7) at (9, 3) {};
		\node [style=none] (8) at (-9, 6) {$p_1$};
		\node [style=none] (9) at (-9, 5) {$p_2$};
		\node [style=none] (10) at (-9, 4) {$p_3$};
		\node [style=none] (11) at (-9, 3) {$p_4$};
		\node [style=VBsmall] (12) at (-5, 5) {};
		\node [style=VBsmall] (13) at (-1, 6) {};
		\node [style=VBsmall] (14) at (5, 6) {};
		\node [style=VBsmall] (15) at (3, 4) {};
		\node [style=VBsmall] (16) at (7, 3) {};
		\node [style=VGsmall] (17) at (-3, 5) {};
		\node [style=VGsmall] (18) at (1, 3) {};
		\node [style=VBsmall] (19) at (-5, 1) {};
		\node [style=VBsmall] (20) at (-2, 1) {};
		\node [style=VBsmall] (21) at (-2, -2) {};
		\node [style=VBsmall] (22) at (-5, -2) {};
		\node [style=VGsmall] (23) at (2, 0) {};
		\node [style=VGsmall] (24) at (5, 0) {};
		\node [style=none] (25) at (-5, 1.75) {$p_1$};
		\node [style=none] (26) at (-2, 1.75) {$p_2$};
		\node [style=none] (27) at (-5, -2.75) {$p_3$};
		\node [style=none] (28) at (-2, -2.75) {$p_4$};
		\node [style=none] (29) at (2, 0.75) {$p_2$};
		\node [style=none] (30) at (5, 0.75) {$p_4$};
		\node [style=none] (31) at (-3.5, -4) {$G_1$};
		\node [style=none] (32) at (3.5, -4) {$G_2$};
	\end{pgfonlayer}
	\begin{pgfonlayer}{edgelayer}
		\draw [style=Eout] (0) to (4.center);
		\draw [style=Eout] (1.center) to (5.center);
		\draw [style=Eout] (2.center) to (6.center);
		\draw [style=Eout] (3.center) to (7.center);
		\draw [style=Eout] (0) to (12);
		\draw [style=Eout] (13) to (15);
		\draw [style=Eout] (14) to (16);
		\draw [style=Eout] (17) to (18);
		\draw [style=Eout] (23) to (24);
		\draw [style=Eout] (19) to (20);
		\draw [style=Eout] (19) to (21);
		\draw [style=Eout] (19) to (22);
	\end{pgfonlayer}
\end{tikzpicture}

%% file: drawings/cycle43.tikz
\begin{tikzpicture}
	\begin{pgfonlayer}{nodelayer}
		\node [style=VBsmall] (0) at (-8, 6) {};
		\node [style=none] (1) at (-8, 5) {};
		\node [style=none] (2) at (-8, 4) {};
		\node [style=none] (3) at (-8, 3) {};
		\node [style=none] (4) at (9, 6) {};
		\node [style=none] (5) at (9, 5) {};
		\node [style=none] (6) at (9, 4) {};
		\node [style=none] (7) at (9, 3) {};
		\node [style=none] (8) at (-9, 6) {$p_1$};
		\node [style=none] (9) at (-9, 5) {$p_2$};
		\node [style=none] (10) at (-9, 4) {$p_3$};
		\node [style=none] (11) at (-9, 3) {$p_4$};
		\node [style=VBsmall] (12) at (-6, 5) {};
		\node [style=VBsmall] (13) at (-5, 5) {};
		\node [style=VBsmall] (14) at (-4, 3) {};
		\node [style=VBsmall] (15) at (-3, 3) {};
		\node [style=VBsmall] (16) at (-1, 4) {};
		\node [style=VBsmall] (17) at (0, 4) {};
		\node [style=VBsmall] (18) at (1, 6) {};
		\node [style=VBsmall] (19) at (2, 6) {};
		\node [style=VBsmall] (20) at (3, 5) {};
		\node [style=VBsmall] (24) at (4, 5) {};
		\node [style=VBsmall] (25) at (5, 3) {};
		\node [style=VBsmall] (26) at (-7, 1) {};
		\node [style=VBsmall] (27) at (-4, 1) {};
		\node [style=VBsmall] (28) at (-4, -2) {};
		\node [style=VBsmall] (29) at (-7, -2) {};
		\node [style=none] (30) at (-7, 1.75) {$p_1$};
		\node [style=none] (31) at (-4, 1.75) {$p_2$};
		\node [style=none] (32) at (-7, -2.75) {$p_3$};
		\node [style=none] (33) at (-4, -2.75) {$p_4$};
		\node [style=VGsmall] (34) at (-6, 3) {};
		\node [style=VGsmall] (35) at (-2, 6) {};
		\node [style=VGsmall] (36) at (1.5, 4) {};
		\node [style=VGsmall] (37) at (2.5, 3) {};
		\node [style=VGsmall] (38) at (5.5, 6) {};
		\node [style=VGsmall] (39) at (3, 1) {};
		\node [style=VGsmall] (40) at (1, -2) {};
		\node [style=VGsmall] (41) at (5, -2) {};
		\node [style=none] (42) at (3, 1.75) {$p_4$};
		\node [style=none] (43) at (5, -2.75) {$p_1$};
		\node [style=none] (44) at (1, -2.75) {$p_3$};
		\node [style=none] (45) at (-5.5, -3.75) {$G_1$};
		\node [style=none] (46) at (3, -3.75) {$G_2$};
	\end{pgfonlayer}
	\begin{pgfonlayer}{edgelayer}
		\draw [style=Eout] (0) to (4.center);
		\draw [style=Eout] (1.center) to (5.center);
		\draw [style=Eout] (2.center) to (6.center);
		\draw [style=Eout] (3.center) to (7.center);
		\draw [style=Eout] (0) to (12);
		\draw [style=Eout] (13) to (14);
		\draw [style=Eout] (15) to (16);
		\draw [style=Eout] (17) to (18);
		\draw [style=Eout] (19) to (20);
		\draw [style=Eout] (24) to (25);
		\draw [style=Eout] (26) to (27);
		\draw [style=Eout] (27) to (28);
		\draw [style=Eout] (28) to (29);
		\draw [style=Eout] (29) to (26);
		\draw [style=Eout] (34) to (35);
		\draw [style=Eout] (35) to (36);
		\draw [style=Eout] (36) to (37);
		\draw [style=Eout] (37) to (38);
		\draw [style=Eout] (39) to (41);
		\draw [style=Eout] (41) to (40);
		\draw [style=Eout] (40) to (39);
	\end{pgfonlayer}
\end{tikzpicture}

%% file: 3-results.tex
\section{The Firing Rebels Problems} \label{sec:results}

In this section, we will apply our communication abstractions
to the \emph{firing rebels with relay} (FRR) problem and its simpler 
variant \emph{firing rebels} (FR), which have been introduced and epistemically analyzed in~\cite{FireTark}.
They assume a purely asynchronous distributed system with $n \geq 2f+1$ processes, up to $f$ of which may be byzantine faulty, where every process may experience an external event $\START$ at most once, at any time. Processes can communicate by exchanging messages, with the goal to eventually execute a special action $\FIRE$ at most once. Informally, for FR, it is required that if enough correct processes observe $\START$ in a run, then all correct processes should eventually execute $\FIRE$. In order to prevent trivial solutions, however, a correct process may execute $\FIRE$ only when $\START$ occurred at some correct process. For FRR, it is also required that if some correct process executes $\FIRE$, all correct processes must eventually do so. Note that every correct implementation of FRR needs $n\geq 3f+1$ processes.

\begin{definition}[Firing rebels problems {\cite[Def.~1]{FireTark}}]\label{def:firing}
	A system is consistent with FR for $f\geq0$ when all runs satisfy:
	\begin{enumerate}
		\item[(C)] \emph{Correctness}: If at least $2f+1$ processes learn\footnote{This term has been used in \cite {FireTark} to be as generic as possible here. In particular, it is satisfied in the case when $2f+1$ processes, correct or not, witness $\START$ locally, as well as in the case where $2f+1$ processes are informed that $f+1$ correct processes have witnessed $\START$ locally.} that $\START$ occurred at a correct process, then all correct processes perform $\FIRE$ eventually.
		\item[(U)] \emph{Unforgeability}: If a correct process executes $\FIRE$, then $\START$ occurred at a correct process.
    \end{enumerate}
Moreover, the system is consistent with FRR if every run also satisfies:
    \begin{enumerate}
  \item[(R)] \emph{Relay}: If a correct process executes $\FIRE$, all correct processes execute $\FIRE$ eventually.
	\end{enumerate}
\end{definition}

Throughout this paper, when the precondition of (C) is not just vacuously satisfied, we will assume runs where $f+1$ correct processes locally observe $\START$, and that there are $f$ byzantine faulty processes that observe $\START$ locally as well. 

We start our considerations by recalling some graph theoretic definitions and lemmas.

\begin{definition}[Strong vertex connectivity]\label{def:vertexconn}
    Let $G = \langle V, E \rangle$ be a directed graph with respective set of vertices $V$ and edges $E$. We say that $G$ is $k$-strongly vertex connected, if $k$ is the minimum integer such that $\vert V \vert > k$ and for any vertex subset $S \subset V$ with $\vert S \vert < k$, the subgraph induced by $V \setminus S$ is strongly connected.
\end{definition}

Although Menger's theorem has originally been stated for undirected graphs, it can also be stated in terms of directed graphs.

\begin{theorem}[Menger's theorem for directed graphs {\cite[Thm.~4.2.17]{IntroGraphTh}}]\label{thm:menger}
    Let $G$ be a directed graph. $G$ is $k$-strongly vertex connected iff for any pair of vertices $v,w$, $v \neq w$, there exist at least $k$ internally disjoint paths, i.e., directed paths that are vertex disjoint
    except for the starting and ending vertex, from $v$ to $w$.
\end{theorem}

The following \cref{def:strroot} will turn out to be instrumental in our analysis:

\begin{definition}[Strong root]\label{def:strroot}
    Let $G$ be a directed graph and $S \subseteq V(G)$. We say that $S$ is a $k$-strong root of $G$ if, for any vertex $v \notin S$, there exist at least $k$ prefix-disjoint paths from $S$ to $v$, i.e., paths that are vertex disjoint except for the common endpoint $v$.
\end{definition}

The following \cref{lem:source} shows that $k$-strong vertex connectivity implies that every vertex is reachable from any source $S$ of size $k$ by prefix-disjoint paths, i.e., that $S$ is a strong root: 

\begin{lemma}[Source Lemma]\label{lem:source}
    Let $G$ be a $k$-strongly vertex connected directed graph. Then, every set of vertices $S$ with $\vert S \vert = k$ is a strong root.
\end{lemma}
\begin{proof}
Consider a set $S$ such that $\vert S \vert = k$ and a vertex $v \notin S$. Let $w$ be a vertex foreign to $G$, i.e. $w \notin V(G)$. Consider graph $G' = \langle V(G'), E(G') \rangle$, where $V(G') = V(G) \cup \{ w\}$ and $E(G') = E(G) \cup \{ (w,u), (u,w) \; \vert \; u \in S\}$. 

We first prove that $G'$ is also $k$-strongly vertex connected. Let $U \subset V(G')$ be a vertex set such that $\vert U \vert < k$. If $w \in U$, then $G' \setminus U = G \setminus (U \setminus \{w\})$. Since $\vert U \setminus \{w\} \vert < \vert U \vert < k$, it immediately follows that $G' \setminus U$ is strongly connected. If $w \notin U$, it is sufficient to show that $w$ is reachable and can reach the rest of $G \setminus U$: Since $\vert U \vert < k$, and $\vert S \vert = k$, we have $S \setminus U \neq \varnothing$. Picking any $u \in S \setminus U$, we must also have $u \in G \setminus U$, and $(w,u)$ and $(u,w) \in G' \setminus U$. Since $u \in G \setminus U$, vertex $u$ is strongly connected to all vertices in $G \setminus U$. Therefore, $w$ is indeed strongly connected to all vertices in $G \setminus U$, which completes our proof that $G'$ is $k$-strongly vertex connected.

From \cref{thm:menger}, it follows that there are at least $k$ internally disjoint paths $P_1, \ldots, P_k$ from $w$ to any other vertex $v \notin S$ in $G'$. Since $w$ is only adjacent to $S$, every $P_i \setminus \{w\}$, $1 \leq i \leq k$, is necessarily a path from some vertex in $S$ to $v$. Since all $P_i$ are internally disjoint, all $P_i \setminus \{w\}$ are prefix-disjoint. Therefore, $S$ is a $k$-strong root.
\end{proof}

The following \cref{thm:connroot} shows that $k$-strong vertex connectedness and $k$-strong rootedness are in fact equivalent:

\begin{lemma}\label{thm:connroot}
    Let $G$ be a directed graph. $G$ is $k$-strongly vertex connected iff every set $S$ of vertices with $\vert S \vert = k$ is a $k$-strong root.
\end{lemma}
\begin{proof}
    Assume that $G$ is $k$-strongly vertex connected, and $S$ such that $\vert S \vert=k$. It follows from \cref{lem:source} that $S$ is a $k$-strong root.

    Now assume that $G$ is not $k$-strongly vertex connected. Then there exists a vertex set $S\subseteq V(G)$ of size $k-1$ that separates a vertex $x$ from a vertex $y$, i.e., $x$ cannot reach $y$ in $G \setminus S$. It follows that $S \cup \{x\}$ cannot be a $k$-strong root of $G$, as $k$ prefix-disjoint paths from $S \cup \{x \}$ to $y$ would exist otherwise. Since $\vert S \vert = k-1$, there is at least one path from $x$ to $y$ that bypasses $S$, contradicting that $S$ separates $x$ from $y$.
\end{proof}

\subsection{Firing Rebels (FR)}

We are now ready to introduce and analyze a simple deterministic protocol for solving FR in a purely asynchronous byzantine system with $n \geq 2f+1$ processes. Informally, it works as follows: Every process $p$ maintains an initially empty collection $H_p$ of message chains, each of which attests to a $\START$ event. For example, $p_1(p_4(\START))$ at process $p$ attests that a $\START$ event has occurred at $p_4$, which has been communicated to $p_1$ and finally to $p$. This communication is implemented by letting every $p$ broadcast its $H_p$ as soon as it is non-empty. Upon receiving $H_q$ from process $q$, process $p$ updates its own $H_p$ by adding all message chains in $H_q$, properly extended by appending $q$, that do not already contain $p$. Once $p$ either witnesses a $\START$ event itself or finds $f+1$ disjoint chains in $H_p$, it executes $\FIRE$ and broadcasts the singleton set $H_p=\{\START\}$ forever after.

\begin{algorithm}[!h]
\caption{A protocol for firing rebels without relay, code for process $p$}\label{alg:FR}
$H \gets \varnothing$\;

\While{$\mathbf{true}$}{
    $(Hq,q) \gets \mathbf{receive}()$ /* receive msg containing $Hq$ from $q\neq p$, empty if none */\;
    \While{$msc \in Hq$}{ /* process all message chains $msc$ in $Hq$ */\;
            \If{$p \notin msc$}{
            $H \gets H \cup \{q(msc)\}$\; 
        }
        $Hq \gets Hq\setminus msc$\;
    }
    \If{$disjoint_{f+1}(H) \vee witness(\START) $}{
        $H \gets \{ \START\}$\;
        $\FIRE$\; \label{line:fire}
        \While{$\mathbf{true}$}
            {
                $\mathbf{broadcast}(H)$\;\label{line:bc2}
            }
    }
    $\mathbf{broadcast}(H)$\;\label{line:bc1}
}
\end{algorithm}

\cref{alg:FR} provides the pseudo-code of this algorithm, for process $p$: \textbf{receive()} delivers a message received from some process, if there is any; $witness(\START)$ tells whether $p$ locally witnessed the external event $\START$, \textbf{broadcast($H$)} sends a message containing $H$ to all other processes, and $disjoint_{f+1}(H)$ tells whether $H$ contains $f+1$ disjoint message chains.

We will now prove that \cref{alg:FR} is correct for every run governed by a communication abstraction $ST \cdot G$ as
defined in \cref{lem:alg1corr}.
 
\begin{lemma}[FR Correctness]\label{lem:alg1corr}
    \cref{alg:FR} satisfies Correctness (C) for every run $r$ governed by a communication abstraction $ST \cdot G$, where $ST$ is a $\START$ event graph with participating set $S_{ST}$ of processes in $ST$ such that either (i) $\vert S_{ST} \vert < 2f+1$ and $G$ is arbitrary or (ii) there is a subset $S \subseteq S_{ST}$ that is a $2f+1$ strong root of $G$.
\end{lemma}

\begin{proof}
 As for case (i), if the $\START$ event is witnessed by less than $2f+1$ processes, (C) holds vacuously. In case (ii), we know that all the $2f+1$ processes in $S$ witness $\START$. Note that all correct processes from $S$, as well as all other processes in $S_{ST}$, are hence guaranteed to execute $\FIRE$ in Line~\ref{line:fire} by our code.
 Consider a correct process $v$ such that $v\notin S$. By \cref{def:strroot}, there are $2f+1$ prefix-disjoint paths from $S$ to $v$ in $G$. Since all paths are prefix-disjoint and there are at most $f$ faulty processes, there are (at least) $f+1$ such paths $P_1, \ldots, P_{f+1}$ in $G$ containing only correct processes.

 Now consider a process $w \in P_i$. Since $w$ is contained in a path in $G$, it has the opportunity to broadcast its $H_w$, scheduled for some time $t_w$. Since $ST \cdot G$ governs our run, there are two possibilities: If $w\in S_{ST}$ and witnesses $\START$ already before $t_w$ in $P_i$, it will execute $\FIRE$ and send $H_w=\{\START\}$ to its successor in $P_i$ at time $t_w$ according to Line~\ref{line:bc2}. Otherwise, $w$ forwards 
 the prefix of $P_i$, received from its predecessor in $P_i$ and thus maintained in $H_w$, to its successor in $P_i$. 
 A trivial induction based on this argument reveals that $v$ will eventually receive a message chain via each of the $f+1$ paths $P_i$. Since they are all prefix-disjoint and formed only by correct processes, $v$ will eventually execute $\FIRE$ in Line~\ref{line:fire}. 
\end{proof}

The following lemma shows that \cref{alg:FR} guarantees (U) in any run.

\begin{lemma}[FR Unforgeability] \label{lem:alg1unfor}
\cref{alg:FR} satisfies Unforgeability (U) for any run $r$, governed by any communication abstraction.
\end{lemma}
\begin{proof}
If no process executes $\FIRE$ in a run, (U) is vacuously true. Otherwise,
    let $p$ be any correct process executing \cref{alg:FR} that executes $\FIRE$. Since Line~\ref{line:fire} is the only line where this can happen, $p$ either witnessed a $\START$ event or it received at least $f+1$ prefix-disjoint message chains attesting $\START$. Since there are at most $f$ faulty processes, at least one of the
    originators of the paths must be a correct process that witnessed $\START$.
\end{proof}

Conversely, the following \cref{lem:frimp} shows that FR cannot be solved, by any protocol, if there are runs that are not governed by a communication abstraction as specified in \cref{lem:alg1corr}. More specifically, FR cannot be solved
if there are runs governed by a communication abstraction $ST \cdot G$ where $\vert S_{ST} \vert \geq 2f+1$ but there is no subset $S \subseteq S_{ST}$ that is a $2f+1$-strong root.

\begin{lemma}[FR impossibility] \label{lem:frimp}
Consider any protocol $P$ in an asynchronous byzantine $f$-resilient system, which aims at solving FR. If a communication abstraction $ST \cdot G$ exists in that system, where the $\START$ event graph $ST$ has a participating set $S_{ST}$ of processes with $\vert S_{ST} \vert \geq 2f+1$ but there is no subset $S\subseteq S_{ST}$ that is a
$2f+1$-strong root, then $P$ does not solve FR in some runs. More specifically, for every choice of a set $S$ of $2f+1$ processes and every set $B\subseteq S$, there is a process
$v \notin S$ that does not execute $\FIRE$ in the run $r'$ governed by $ST \cdot G$, where
the processes in $B$ are faulty in that they pretend that they did not witness $\START$ and behave accordingly like a correct process would do.
\end{lemma}

\begin{proof}
Suppose, for a contradiction, that there is some protocol that solves FR despite the fact that a communication
abstraction $ST \cdot G$ exists in the system, where $\vert S_{ST} \vert \geq 2f+1$ but no subset 
$S\subseteq S_{ST}$ with $\vert S \vert = 2f+1$ is a $2f+1$-strong root in $r$. For every such set $S$, \cref{def:strroot} implies that there is some process $v \notin S$ that is not reachable from $S$ by $2f+1$ prefix-disjoint paths in $G$.
Let $\mathcal{P}$ be a maximal set of prefix-disjoint paths from $S$ to $v$, where of course $\vert \mathcal{P} \vert < 2f+1$. Let us first consider the case where $\vert \mathcal{P} \vert > f$. Let $B \subseteq S$ be a set of $f$ different starting vertices of paths in $\mathcal{P}$. Otherwise, if $\vert \mathcal{P} \vert \leq f$, then we define $B$ as the starting vertices in $\mathcal{P}$ and sufficiently many other initial vertices in $S$. In any case, $S \setminus B$ must have less than $f+1$ vertex-disjoint paths to $v$. Therefore, there is a set $U$ of at most $f$ vertices that break each path from $S \setminus B$ to $v$. 
Clearly, in a run $r$ adhering to $ST \cdot G$ where no process is byzantine faulty, $v$ must execute $\FIRE$ eventually in $r$ by Correctness (C). 
    
Now consider a run $r'$ governed by the communication abstraction $ST \cdot G$, where the processes in 
$B$ are byzantine faulty: they pretend that they did not witness a $\START$ event, albeit they did so, and behave exactly as a correct process that did not witness $\START$ otherwise. Note carefully that such a run $r'$ must indeed exist, since our adversary has been assumed to be fault-oblivious, recall \cref{def:adversary}. Albeit the runs $r$ and $r'$ are not indistinguishable for $v$, Correctness (C) requires
that $v$ must execute $\FIRE$ eventually in $r'$ as well.

Finally, consider another communication abstraction $ST''\cdot G$, where no $\START$ event ever happens, i.e., where the participating set of processes is $S_{ST''}=\emptyset$. Consider the run $r''$ adhering to $ST''\cdot G$, where only the processes 
in $U$ are are byzantine faulty: they simulate the behavior that they exhibited in run $r$ w.r.t.\ the paths from $S \setminus B$ to $v$. By construction, the runs $r'$ and $r''$ are indistinguishable for $v$, so $v$ should execute $\FIRE$ eventually in $r''$ again. However, since $S_{ST''}=\emptyset$, no process and hence no correct
process witnessed $\START$, so that $v$ must not execute $\FIRE$ by Unforgeability (U). This provides the required contradiction.

The run $r'$ stated in our lemma is just the one used in the above construction: Since we just proved that $v$ does not execute $\FIRE$ in $r''$, it cannot execute $\FIRE$ in $r'$ due to indistinguishability either.
\end{proof}

\begin{theorem}[FR solvability] \label{thm:firingrebels}
    FR is solvable in a byzantine asynchronous system with $n\geq 2f+1$ processes, iff every run is governed by
    a communication abstraction $ST \cdot G$ comprising a $\START$ event graph $ST$ with participating set $S_{ST}$ of processes in $ST$ such that either (i) $\vert S_{ST} \vert < 2f+1$ and $G$ is arbitrary or (ii) there is a subset $S \subseteq S_{ST}$ that is a $2f+1$ strong root of $G$.
\end{theorem}

\begin{proof}
    Solvability is guaranteed for \cref{alg:FR} by \cref{{lem:alg1corr}} and \cref{{lem:alg1unfor}}. The matching impossibility is given by \cref{lem:frimp}.
\end{proof}

\subsection{Firing Rebels with Relay (FRR)}
\label{sec:FRR}

We now turn our attention to the FRR problem, which, unlike FR, is of substantial practical interest as a building block in several distributed algorithms, recall \cref{sec:intro}. In addition to Correctness (C) and Unforgeability (U), it also needs to satisfy the Relay (R) property in \cref{def:firing}. Note carefully that (R) is an all-or-nothing and hence an agreement-type property. Since (C) in conjunction with (R) imply that a quorum of processes witnessing $\START$, which is sufficiently large to cause just one agent to execute $\FIRE$, must eventually cause \emph{all} correct agents to do so as well, it is obvious that they (also) involve a liveness property. It is hence natural to suspect that \emph{finite} network abstractions are insufficient for correctly solving FRR in byzantine asynchronous systems, and we will prove this conjecture to be true.

In order to characterize the solvability of FRR, we need to introduce another graph-theoretic concept similar to strong roots. Recall that the latter denotes a subset of vertices that \emph{can reach} every vertex in the graph via sufficiently many prefix-disjoint
paths. Now we are interested in a set of vertices that \emph{can be reached} via sufficiently many prefix-disjoint paths, which we will call a co-root.

\begin{definition}\label{def:coroot}
    Let $G$ be a directed graph. A vertex set $C$ is a $k$-co-root of $G$ with respect to a vertex set $B$, if for every vertex $v \in C$, there is a vertex set $S(v)$ that can reach $v$ through $k$ prefix-disjoint paths that do not intersect $B$.
\end{definition}

\begin{definition}\label{def:res-corooted}
    Let $G$ be a directed graph. We say that $G$ is $(k,l,m)$ co-rooted if for any vertex set of size $m$, there exists a $k$ co-root of size $l$.
\end{definition}

\begin{lemma}\label{lem:degreecond}
$C$ is a $k$-co-root of a directed graph $G$, iff the minimum in-degree of all the vertices in $C$ in the graph $G \setminus B$, denoted by $\delta_{G \setminus B}^-(C)$, is at least $k-1$.
\end{lemma}
\begin{proof}
    The only-if direction follows from the fact that if a vertex $v$ has an in-degree less than $k$, then it can not be reached by $k$ prefix-disjoint paths. Conversely, since $G$ is a simple graph and hence has at most one directed edge per vertex pair $(w,v)$, the incoming edges of vertex $v$ with in-degree $k-1$ form $k-1$ prefix-disjoint paths. The missing remaining path is the path of length 0 consisting of only vertex $v$. This path is vacuously prefix-disjoint from all the rest of the incoming edges.
\end{proof}

Note carefully that whereas every set of $2f+1$ processes is a $2f+1$-strong root in a $2f+1$-strongly vertex connected graph $G$ by \cref{lem:source}, it does not necessarily contain a $2f+1$-co-root. To see this, just take the fully connected graph $G$ with 
$2f+2$ vertices. If one removes a set $B$ of size $f$, the in-degree of the remaining
vertices is only $f+1$. Hence, \cref{lem:degreecond} reveals that no $2f+1$-co-root with
respect to $B$ can exist.

\medskip

We start our considerations with a solution algorithm, which solves FRR in an asynchronous byzantine system with $n \geq 3f+1$ processes. It has been inspired by the consistent broadcasting primitive from \cite{ST87}, but non-trivially improves it in terms of network connectivity requirements. Our algorithm essentially employs two concurrent instances of the FR \cref{alg:FR}: The first instance uses message chains attesting to $\START$, like the original algorithm. However, rather than triggering $\FIRE$ when $f+1$ disjoint message chains have been observed, it requires $2f+1$ disjoint message chains to do so. The second instance uses message chains attesting to $\FIRE$, and can also trigger the execution of $\FIRE$, namely, when $f+1$ disjoint $\FIRE$ message chains have been observed. Note that it is this
use of another type of message chains that distinguishes our algorithm from implementations such as \cite{WS09:DC,RS11:TCS}, which
need a fully connected network, i.e., network abstractions with $3f+1$ vertex connectivity.

The pseudo-code of our FRR algorithm, for process $p$, is given in \cref{alg:FRR}. Like in FR, \textbf{receive()} delivers a message received from some process, if there is any; $witness(\START)$ tells whether $p$ locally witnessed the external event $\START$, \textbf{broadcast($M$)} sends a message containing $M \in \{H,V\}$ to all other processes, and $disjoint_{x}(M)$ tells whether $M$ contains $x$ disjoint message chains. $H$ resp.\ $V$ contains all the message chains for $\START$ resp.\ $\FIRE$ that $p$ has seen so far, initially set to empty. The boolean flag $wit$ is set when $p$ has either witnessed $\START$ locally or has seen $f+1$ prefix-disjoint $\START$ message chains for the first time. The boolean flag $done$ is set when $\FIRE$ is executed.

        

\begin{algorithm}[ht]
\caption{A protocol for firing rebels with relay, code for process $p$}\label{alg:FRR}
$H \gets \varnothing$, $V \gets \varnothing$\;
$wit \gets \mathbf{false}$, $done \gets \mathbf{false}$\;
\While{$\mathbf{true}$}{
    $(Hq,Vq,q) \gets \mathbf{receive}()$ /* receive msg containing $Hq$ and $V_q$ from $q\neq p$, both empty if none */\;
    \While{$msc \in Hq$}{  /* process all $\START$ message chains $msc$ in $Hq$ */\;
        \If{$p \notin msc$}{
            $H \gets H \cup \{q(msc)\}$\; 
        }
        $Hq \gets Hq\setminus msc$\;
    }
    \While{$msc \in Vq$}{/* process all $\FIRE$ message chains $msc$ in $Vq$ */\;
        \If{$p \notin msc$}{
            $V \gets V \cup \{q(msc)\}$\; 
        }
        $Vq \gets Vq\setminus msc$\;
    }
    \If{$\neg wit \wedge (disjoint_{f+1}(H) \vee witness(\START)) $}{
        $wit \gets  \mathbf{true}$\;\label{line:frrwit}
        $H \gets H \cup \{\START\}$ /* (Pretend that) $\START$ happened locally, but keep the previous content of $H$ for Line~\ref{line:firecond} */
    }
    \If{$\neg done \wedge(disjoint_{f+1}(V) \vee disjoint_{2f+1}(H)) $}{\label{line:firecond}
        $done \gets  \mathbf{true}$\;\label{line:frrdone}
        $H \gets \{\START\}$\;
        $V \gets \{\FIRE\}$\;
        $\FIRE$\;\label{line:frrfire}
        \While{$\mathbf{true}$}{
            $\textbf{broadcast}(H,V)$\;\label{line:frrbc1}
        }
    }
    \If{$wit$}{
        $\mathbf{broadcast}(\{\START\},V)$ /* No need to re-broadcast entire $H$ here */\;
    }
    \Else{
    $\mathbf{broadcast}(H,V)$\;\label{line:frrbc2}
    }
}
\end{algorithm}

We proceed with proving the three properties of FRR stated 
in \cref{def:firing}, for all runs corresponding to certain
communication abstractions.

\begin{lemma}\label{lem:alg2unforg}
     $\cref{alg:FRR}$ satisfies Unforgeability (U) for any run $r$, governed by any communication abstraction.
\end{lemma}

\begin{proof}
    If no process executes $\FIRE$ in a run, (U) is vacuously true.
    Otherwise, assume that some correct process $p$ executes $\FIRE$ in Line~\ref{line:frrfire}. There are two cases: (i) If $H_p$ contains $2f+1$ disjoint $\START$ chains, at least $f+1$ of those must originate in a correct process that has observed $\START$. (ii) If $V_p$ contains $f+1$ disjoint $\FIRE$ chains, at least one of those must originate in a correct process $r$ that executed $\FIRE$. For $r$, however, case (i) must apply. Consequently, some correct process must have locally witnesses $\START$ in any case, as needed by (U).
\end{proof}

\begin{lemma}\label{lem:alg2corr}
    \cref{alg:FRR} satisfies Correctness (C) for any run $r$ governed by a communication abstraction $ST\cdot G_1 \cdot  G_2 \cdot G_3 $, where $ST$ is a $\START$ event graph with participating set $S_{ST}$ such that either (i) $\vert S_{ST} \vert < 2f+1$ and $G_1$, $G_2$ and $G_3$ are arbitrary, or (ii) there is a subset $S \subseteq S_{ST}$ that is a $2f+1$-strong root of $G_1$, $G_2$ has a $2f+1$-co-root $Q$ where $|Q|\geq f+1$ with respect to any set $B$ of size $f$, and $G_3$ is $2f+1$-strongly vertex connected.
\end{lemma}

\begin{proof}
    For case (i), if the $\START$ event is witnessed by less than $2f+1$ processes, (C) holds vacuously. In case (ii), we know that all the $2f+1$ processes in $S$ witness $\START$. Since $ST \cdot G_1$ is the communication abstraction governing $r$ and $S$ is a $2f+1$-strong root of $G_1$, in run $r$, every correct process $p \notin S$ will eventually get at least $f+1$ prefix-disjoint paths from correct processes in $S$ that started after the respective $\START$ events in $S$, since there are at most $f$ byzantine processes. Therefore, $p$ will set $wit$ in Line~\ref{line:frrwit} and pretend that a $\START$ event happened at $p$. On the other hand, every correct process $s \in S$ witnesses $\START$ locally and hence also sets $wit$. In both cases, $p$ initiates a $\START$ message chain by broadcasting $\START$ as part of $H_p$.
    
    For every correct $p \notin S$, let $P_{p}$ be the set of the last vertices of the $f+1$ prefix-disjoint message chains that actually reached process $p$ in $r$. Moreover, let $B$ be the set of $f$ processes that contains all the faulty processes in $r$. Let $Q$, $|Q|\geq f+1$, be a $2f+1$-co-root of $G_2$ with respect to $B$ and consider a arbitrary $q \in Q$. According to \cref{def:coroot} and our choice of $B$, $q$ is necessarily correct and has at least $2f+1$ correct in-neighbors in $G_2$, namely $w_1, \ldots, w_{2f+1}$. Note that each of the edges $(w_1,q), \ldots, (w_{2f+1},q)$ is a path in $G_2$. 
    
    Since $ST \cdot G_1 \cdot G_2$ is the communication abstraction governing $r$, and $(v,w_i) \in G_1$ for every $v\in P_{w_i}$ where $w_i \notin S$, there is a message relay opportunity that is represented by the path $P_{w_i} + (w_i,q)$. For $w_i \in S$, obviously $w_i \in G_1$. 
    Since each $w_i$, as a correct process, broadcasts $\START$ in $r$ as established above, $q$ will receive at least $2f+1$ disjoint $\START$ messages from its in-neighbors. Therefore, $q$ will enter Line~\ref{line:frrdone} and execute $\FIRE$ in Line~\ref{line:frrfire}. Since $q \in Q$ has been chosen arbitrarily, it follows that all the $2f+1$ processes in $Q$ eventually execute $\FIRE$.

    Hence, it only remains to show that all the correct processes outside of $Q$ will also
    execute $\FIRE$ eventually. This is taken care of by the second instance of FR embedded in \cref{alg:FRR}, which utilizes the $\FIRE$ message chains and can also cause the execution of $\FIRE$ in Line~\ref{line:frrfire}. In fact, the part of 
    the run $r$ governed by $ST \cdot G_1 \cdot G_2$, which caused the at least $f+1$ correct
    processes in $Q$ to eventually execute $\FIRE$ and also started corresponding $\FIRE$ 
    message chains, can also be viewed as the generation of ``external events'' $\FIRE$ governed by some appropriate event graph $FT$ with participating set $Q \cup B$ of size $\vert Q \cup B \vert \geq 2f+1$. It is the network abstraction $G_3$ that governs the evolution of these message chains. According to \cref{thm:connroot}, it is $2f+1$-strong vertex connected, hence also guarantees a $2f+1$-strong root. Consequently, $FT \cdot G_3$ can be viewed as the communication abstraction for this second instance of FR. \cref{lem:alg1corr} thus guarantees that all processes outside $Q$ will also execute $\FIRE$ eventually (when they have not done so earlier).
    
    \end{proof}

\begin{lemma}\label{lem:alg2rel}
    \cref{alg:FRR} satisfies Relay (R) for any run $r$ governed by a dynamic network abstraction $\mathcal{G}=G_1 \cdot G_2 \cdots G_k \cdots$, where each $G_i$ is $2f+1$-strongly vertex connected and, for any vertex set $B$ of $f$ processes, has a $2f+1$-co-root $C_i$ with $|C_i|\geq f+1$ with respect to $B$.
\end{lemma}

\begin{proof}
    Assume that a correct process $p$ executes $\FIRE$ in Line~\ref{line:frrfire}. It follows from~\cref{alg:FRR} that either (i) $p$ received $f+1$ prefix-disjoint message chains attesting to a $\FIRE$ message, or (ii) $p$ received $2f+1$ prefix-disjoint message chains attesting to $\START$. In case (i), the originator of at least one of the $\FIRE$ message chains must be a correct process that executed $\FIRE$ according to case (ii). Hence, it suffices to deal with case (ii). 
    
    Clearly, at least $f+1$ of the $2f+1$ message chains received by $p$ must originate in a correct process that locally
    witnessed $\START$. Let $E$ be the set of these processes, and $B$ be a set of $f$ other (faulty) processes. It follows that the communication abstraction that
    governs $r$ starts with an event graph $ST$ for $\START$ with participating set $S_{ST}= E \cup B$ of size $\vert E \cup B \vert \geq 2f+1$.
    
    Our assumptions on the properties of every $G_i$ imply that it also satisfies the conditions stated for any of $G_1$, $G_2$ and $G_3$ in \cref{lem:alg2corr}. Since we ruled out Zeno behavior in run $r$ and $\mathcal{G}$ is infinite, communication
    abstraction $ST \cdot G_i \cdot G_{i+1} \cdot G_{i+1}$ for any suffix of $\mathcal{G}$ satisfies the conditions of \cref{lem:alg2corr}. It hence follows from \cref{lem:alg2corr} that any correct process will eventually execute $\FIRE$.
\end{proof}

We will now turn our attention to a matching impossibility result for FRR. Not surprisingly,
it is the Relay condition (R) that makes it considerably harder to solve FRR. Indeed, whereas the case where less than $2f+1$ processes locally witnessed a $\START$ event was trivial for FR, since (C) does not force a correct process to fire, this is no longer true for FRR: Byzantine processes may cause some correct processes to execute $\FIRE$ also in this case, which in turn would require all correct processes to do the same.

First, we will show that finite communication abstractions are not enough for solving FRR. This implies that
a network that provides a ``reasonable'' connectivity only for a finite period of time is not 
sufficient for solving this problem.

\begin{lemma}\label{lem:infinity}
    Let $ST \cdot G_1 \cdot G_2 \cdots G_k$ be a finite communication abstraction, and $P$ a protocol that correctly solves FR, i.e., satisfies Correctness (C) and Unforgeability (U) in any run governed by $ST \cdot G_1 \cdots G_k$, for any choice of $ST$. Then, at least one of these runs does not solve FRR, i.e., violates (R).
\end{lemma}
\begin{proof}
Assume for a contradiction that there exists a finite network abstraction $G_1 \cdots G_k$ such that $P$ solves FRR in all runs corresponding to $ST \cdot G_1 \cdots G_k$, for all choices of $ST$. Consider a run $r$ governed by this communication abstraction, where the  $\START$ event graph $ST$ has the participating set $S_{ST}=\{p\}$ and all processes are correct. Since $G_1 \cdots G_k$ is a finite network abstraction, there is a finite time $t_1$ such that the suffix of $r$ after $t_1$ is not governed by $ST \cdot G_1 \cdots G_k$ anymore, recall \cref{rem:runsuffix}, i.e., the processes cannot successfully communicate with each other after time $t_1$.

Note that, since $p$ is the only witness to a $\START$ event in run $r$, this run is indistinguishable, for all processes except $p$, from an alternative run $r_1$ that is also
governed by $ST \cdot G_1 \cdots G_k$, where $p$ is the only byzantine process and just simulates being a witness to a $\START$ event. Note that this can be guaranteed since the adversary is fault-oblivious, recall \cref{def:adversary}. Since $P$ solves FR and hence satisfies (U), no process (except that faulty $p$) is allowed to execute $\FIRE$ in $r_1$. Since $r_1$ and $r$ are indistinguishable for every process except $p$, none of these processes can execute $\FIRE$ in $r$ as well.

Now consider a run $r'$, which corresponds to $ST \cdot G_1 \cdots G_k \cdot ST' \cdot G' \cdot G''$, for the graphs $ST'$, $G'$ and $G''$ defined in the sequel.
In $r'$, the processes in the set $B= \{b_1, \ldots , b_f\}$, with $p \notin B$, are byzantine faulty. Still, all processes in the system behave identically in $r'$ and in $r$ until time $t_1$. After time $t_1$, every $b_i \in B$ simulates to witness a $\START$ event and sends messages only to a set $D = \{d_1, \ldots, d_f\}$ of correct processes, where $p \notin D$ as well. 
The corresponding communication abstraction is provided by the $\START$ event graph $ST'$ with participating set $S_{ST'}=B$, and the static network abstraction $G'$, which contains edges $(b,d)$ for all $b\in B$ and $d\in D$. Similarly, $G''$ allows all the processes
in $D \cup \{p\}$ to communicate directly with each other. Like for $r$, however, there is a time $t_2$, after which the processes cannot communicate with each other in $r'$ anymore.

Now consider a run $r_2$ that corresponds to $ST \cdot G_1 \cdots G_k \cdot ST'' \cdot G' \cdot G'' \cdot G'''$, where the $\START$ participating set of $ST''$ is $S_{ST''}=D \cup B$. In $r_2$, we assume that all processes are correct and behave exactly like in $r'$ until time $t_2$. After $t_2$, the graph $G'''$ allows the processes in $D \cup B \cup \{ p\}$ to send messages to the rest of the processes $\Pi \setminus (D\cup B \cup \{p\})$.

For the processes in $D \cup \{p \}$, the runs $r'$ and $r_2$ are indistinguishable. Since $P$ solves FR and hence satisfies (C), all correct processes must eventually execute $\FIRE$ in $r_2$. Consequently, in $r'$, the processes in $D \cup \{p \}$ must eventually execute $\FIRE$ as well. On the other hand, for the processes in $\Pi \setminus (D\cup B \cup \{p\})$, (i) $r'$ is indistinguishable from $r$ until time $t_1$, and (ii) none of these processes receives any message in $r'$ after $t_1$. Since no correct process (except possibly $p$) fires in run $r$, none of the processes in $\Pi \setminus (D\cup B \cup \{p\})$ executes $\FIRE$ in $r'$ either. Therefore, (R) does not hold in $r'$.
\end{proof}

Since \cref{lem:infinity} implies that finite communication abstractions are not
sufficient for solving FRR, we will turn our attention to dynamic communication 
abstractions.


\begin{lemma}\label{lem:infconnected}
Let $\mathcal{G}= G_1 \cdot G_2 \cdots G_i \cdots$ be a dynamic network abstraction, where only finitely many of the graphs $G_i$ are $2f+1$-strongly vertex-connected, and $P$ be a protocol that correctly solves FR, i.e., satisfies Correctness (C) and Unforgeability (U) in every run governed by every communication abstraction $\mathcal{G}'$ involving $\mathcal{G}$. Then, at least one of these runs does not solve FRR, i.e., violates (R).
\end{lemma}
\begin{proof}

Consider a run $r$ governed by the communication abstraction $\mathcal{G}'=ST \cdot \mathcal{G}$, where the $\START$ event graph has the participating set $S_{ST} = \{p\}$, $p \notin Z$. 
According to our assumptions, there is a time $t_0$ after which some set $Z$ of $2f+1$ processes is never a $2f+1$-strong root in the network abstraction $G_k \cdot G_{k+1} \cdots$
governing the suffix of $r$ after $t_0$. 

Now consider the communication abstraction $ST \cdot G_1 \cdots G_k \cdot ST' \cdot G_{k+1} \cdots$, which has an event graph $ST'$ with participating set $S_{ST'}=B=\{b_1,\ldots, b_f\} \subseteq Z$ inserted in the communication abstraction for $r$. Consider the corresponding run $r'$, where all processes in $B$ are byzantine faulty: Up to time $t_0$, they all behave exactly like a correct process. After time $t_0$, every $b_i \in B$ simulates to witness a $\START$ event, and subsequently behaves like a correct process that has indeed witnessed $\START$. 
Again, this can be guaranteed since the adversary is fault-oblivious, recall \cref{def:adversary}.

Moreover, consider the communication abstraction $G_1 \cdots G_k \cdot ST'' \cdot G_{k+1}' \cdot G_{k+2} \cdots$, which almost the same network abstraction as the one for $r'$. The difference is that it involves $ST''=ST \cup ST' \cup B'$ for a set of processes $B' =\{b_1', \ldots, b_f'\}\subseteq Z \setminus B$, and the graph $G_{k+1}'=G_{k+1} \cup G'$, where $G''$ contains all edges $(z,w)$ for $z \in Z$ and a fixed process $w \in \Pi \setminus Z$. Note that $G_{k+1}'$ still does not contain any $2f+1$-strong root, but makes sure
that the process $w$ does have $2f+1$ prefix-disjoint paths from $Z$ to it. Now take the corresponding run $r_2$, where all processes except the ones in $B'$ are correct; note that the $\START$ events of the processes in $B$ (which were simulated by the faulty processes in $r'$) are now authentic.
The behavior of the faulty processes in $r_2$ is again such that they perform exactly like a correct process up to time $t_0$, then simulate a fake $\START$ event, and subsequently behave like a correct process that has indeed witnesses $\START$. Since the participating set of $ST''$ has size $\vert S_{ST''} \vert =2f+1$ and $G_{g+1}$ does not have a $2f+1$-strong root,
\cref{lem:frimp} reveals that $P$ cannot solve FR in the run $r_2'$, which is identical to 
$r_2$ except that the processes in $B'$ are also correct: There is a correct process $v \notin Z$ that does not execute $\FIRE$. Since $r_2'$ and $r_2$ are indistinguishable for all processes except the ones in $B' \subseteq Z$, $v$ does not execute $\FIRE$ in $r_2$ either.

Finally, consider a the communication abstraction $G_1 \cdots G_k \cdot ST'' \cdot G_{k+1}'' \cdot G_{k+2}'' \cdots$, where $G_{k+j}''=G_{k+j} \cup G''$, $j\geq 1$, and $G''$
contains all edges $(z,w)$ for $z \in Z$ and $w \in \Pi \setminus Z$. Note that $Z$ 
is now a $2f+1$-strong root in every $G_{k+j}''$, and that the same paths end in $w$ 
in both $G_{k+1}'$ and $G_{k+1}''$. Taking the run $r_3$ 
governed by this communication abstraction, where all processes are correct, it follows
from the assumption that $P$ correctly solves FR, hence respects (C), and $\vert S_{ST''} \vert =2f+1$ that all correct processes, and hence also $v$ and $w$, must execute $\FIRE$ in $r_3$. 

By construction, $r_2$ and $r_3$ are indistinguishable for both $v$ and $w$. This implies
that, in $r_2$, process $w$ executes $\FIRE$ but $v$ does not. Since both are correct in
$r_2$, (R) is not satisfied in this run, as asserted in our lemma.
\end{proof}

\begin{remark}\label{rem:strengthening}
Note that if $r$ admits a dynamic network abstraction where every set $Z$ is a $2f+1$ strong root infinitely often, then $r$ also admits a dynamic network abstraction where all the graphs are $2f+1$ strongly vertex-connected. Simply notice that for each vertex set $U$ of size $2f+1$, there is a graph $G_U$ in the dynamic network abstraction that repeats infinitely often, and has $U$ as a $2f+1$ strong root. It follows from \cref{rem:graphcompressor} and \cref{thm:connroot} that all of the $G_U$ graphs can be compacted into a $2f+1$ strongly vertex connected graph that repeats infinitely often. The result of \cref{lem:infconnected}
hence also holds for this relaxed communication abstraction.
\end{remark}

As the final ingredient for the proof of our main result \cref{thm:firingrebelsrelay}, we
introduce a class of FRR protocols called \emph{quorum-based protocols}. In essence, an $f+k$-quorum-based protocol has runs where some correct process executes $\FIRE$ already when $k$ correct processes locally observed $\START$; note that $k$ may of course depend on the failure bound $f$ here. Note that $1 \leq k \leq f+1$ for any correct FRR protocol, as (C) requires all correct processes to execute $\FIRE$ when
$k=f+1$, whereas (U) forbids this when $k=0$. Surprisingly, it will turn out that \emph{every} correct FRR protocol is actually a $2f+1$-quorum-based protocol.

\begin{definition}[Quorum-based protocols]\label{def:quorum}
Consider the runs of an FRR protocol $P$ in the special case where a set $B$ of $f$ processes are byzantine faulty and a set $C$ of $k$ correct processes locally observed $\START$. Every byzantine process pretends to have locally observed $\START$ as well, but behaves like a correct process that did not witness $\START$ towards the processes in $\Pi \setminus C$, whereas it behaves like a correct process that did witness $\START$ towards the processes in $C$. We call $P$ an $k+f$-quorum-based FRR protocol, if $k$ is the smallest value that causes some correct process to execute $\FIRE$ in any such run.
\end{definition}

\begin{lemma} \label{lem:quorum} Let $P$ be an $k+f$-quorum-based protocol $P$ that satisfies Unforgeability (U). If $f<k<2f+1$, then $P$ does not solve FRR.
\end{lemma}
\begin{proof}
 Assume, for a contradiction, that $P$ does solve FRR in all runs.
 Consider a run $r$ according to \cref{def:quorum} where some correct process, and hence, by (R), all correct processes execute $\FIRE$. Let $v \in \Pi\setminus (B \cup C)$ be any correct process. 

Now consider another run $r'$, where all the processes except the $k$ ones in $C$ are correct. No process in $\Pi\setminus C$ locally perceives $\START$, whereas all faulty processes in $C$ behave exactly as they behaved in run $r$. Since no correct process witnessed $\START$ in $r'$, by (U), no correct process, including $v$, is allowed to execute $\FIRE$. 

However, $r$ and $r'$ are indistinguishable for $v$, which provides the required contradiction.
\end{proof}

\cref{lem:quorum} in conjunction with (C) hence implies that every correct FRR protocol is a $2f+1$-quorum-based protocol.

\begin{lemma}\label{lem:coroot}
Let $\mathcal{G}= G_1 \cdot G_2 \cdots G_i \cdots$ be a dynamic network abstraction, where every $G_i$ is $2f+1$-strongly vertex connected but, for any set $B$ of $f$ processes, at most finitely many $G_i$ have a $2f+1$-co-root $S_i$ of size $f+1$ with respect to $B$. Then, no protocol $P$ can correctly solve FRR in all runs governed by a communication abstraction based on $\mathcal{G}$.
\end{lemma}

\begin{proof}
    
The most crucial consequence of \cref{def:quorum} is that, in any correct FRR protocol including $P$, which necessarily is a $2f+1$-quorum-based protocol, no correct process can execute $\FIRE$ when it only learned about the occurrence of at most $2f$ $\START$ events. Otherwise,
the processes in $C$ would execute $\FIRE$ in the runs described in \cref{def:quorum}. 

Consequently, every process $p$ must eventually learn 
that at least $2f+1$ $\START$ events occurred in the system before it execute $\FIRE$. Process $p$ can learn this either (i) directly, 
by getting at least $2f+1$ prefix-disjoint chains attesting $\START$, or (ii) indirectly, by receiving sufficiently many prefix-disjoint message chains (like the at least $f+1$ $\FIRE$ chains used in \cref{alg:FRR}) attesting that (i) has happened for sufficiently many other processes. Inductively, there must always be a sufficiently large set $I$ of processes that learn via (i) before (ii) can occur. Obviously, in the presence of up to $f$ byzantine faulty processes, $|I| \geq f+1$ is mandatory for avoiding information solely produced by faulty processes to propagate via (ii). On the other hand, each of the processes contained in
$I$ must have been reachable from at least $2f+1$ processes in order to fulfill the requirement in (i).
 
Now, according to the assumptions in \cref{lem:coroot}, there is a time $t_0$ after which no graph occurring in the network abstraction $G_k \cdot G_{k+1} \cdots$, which governs the suffix of $r$ after $t_0$, contains any $2f+1$-co-root $S_i$ of size $f+1$ with respect to any $B$. Let $G_1\cdot G_2 \cdots G_k \cdot ST \cdot G_{k+1}\cdot G_{k+2} \cdots$, where the $\START$ event graph has
a participating set $S_{ST}$ of $2f+1$ correct processes, and consider a run $r$ with $f$ byzantine faulty processes in $B$, which just
remain silent to everybody. Since $P$ guarantees (C), all correct processes eventually execute $\FIRE$, so must have learned via (i) or (ii) that $2f+1$ $\START$ events have occurred. If all correct processes learned this directly via (i), then they must have received $2f+1$ prefix-disjoint chains attesting $\START$. Since all processes in $B$ are silent, this implies that the set of correct processes forms a $2f+1$ co-root w.r.t. $B$ of size at least $f+1$, which is a contradiction. 

On the other hand, if some correct process learned this indirectly via (ii), it must have received at least $f+1$ prefix-disjoint message chains attesting to the existence of $2f+1$ $\START$ events. Since all processes in $B$ are silent, again each of these $f+1$ disjoint chains necessarily originate in a correct process. Inductively, there must hence be a set $I$ of $f+1$ correct processes that learned about the $2f+1$ $\START$ events via (i). It follows that $I$ is again a $2f+1$ co-root w.r.t $B$ of size $f+1$, which provides the required contradiction also in this case.

Therefore, $P$ does not correctly solve FRR in all runs.
\end{proof}

\begin{theorem}[FRR solvability] \label{thm:firingrebelsrelay}
    FRR is solvable in a byzantine asynchronous distributed system with $n\geq 3f+1$ processes, iff any run $r$ is governed by a communication abstraction based on a dynamic network abstraction $\mathcal{G}=G^{\omega}$ where $G$ is a $2f+1$-strongly vertex-connected graph that is also $(2f+1,f+1,f)$-co-rooted.
\end{theorem}

\begin{proof}
Our theorem follows from combining the possibility results \cref{lem:alg2unforg}, \cref{lem:alg2corr} and \cref{lem:alg2rel} with the matching impossibility result \cref{lem:coroot}, in conjunction with the fact that there are only finitely many possible graphs: Since we consider infinite graph sequences in \cref{lem:coroot}, there is a single graph that repeats infinitely often. Moreover, as $2f+1$ connectivity and co-rootedness are closed with respect to union, we can apply \cref{rem:graphcompressor} to obtain a single graph that has all the desired properties and repeats infinitely often.
\end{proof}

We conclude this section by the following \cref{cor:maximal}, which shows that the dynamic network 
abstraction $\mathcal{G}=G^{\omega}$ that is guaranteed by \cref{thm:firingrebelsrelay} to govern 
any correct run of a protocol that solves FRR can even be assumed to be maximal:

\begin{corollary}\label{cor:maximal}
If FRR is solvable in a byzantine asynchronous distributed system with $n\geq 3f+1$ processes, then the graph $G$ in the dynamic network abstraction $\mathcal{G}=G^{\omega}$ guaranteed by \cref{thm:firingrebelsrelay} is not only $2f+1$-strongly vertex-connected and $(2f+1,f+1,f)$-co-rooted, but can also be assumed to be eventually maximal, in the sense that, eventually, successful communication is only
possible between processes connected by an edge in $G$.
\end{corollary}

\begin{proof}
    Since FRR is solvable by assumption, \cref{thm:firingrebelsrelay} ensures that any run $r$ is governed
    by a network abstraction $G^{\omega}$ where $G$ is $2f+1$-strongly vertex-connected and $(2f+1,f+1,f)$-co-rooted. Consider the directed graph $G'$ with $V(G_{\sigma}')= \Pi$ and $E(G) = \{(p,q) \; \vert \; \mbox{$p$ can successfully send messages to $q$ infinitely often in $r$}\}$. Since $(G)^{\omega}$ is dynamic network abstraction governing $r$, it follows immediately from the definition that $G \subseteq G'$. Therefore $G'$ is $2f+1$-strongly vertex-connected and $(2f+1,f+1,f)$-co-rooted as well. From the definition of $G'$, it follows that, after some finite time $t$, messages can successfully transmitted only via the edges of $G'$. Therefore, 
    $(G')^{\omega}$ is also a dynamic network abstraction governing $r$, which is eventually maximal.    
\end{proof}

An interesting consequence of \cref{cor:maximal} is that the graph $G$ that defines the eventually maximal dynamic network abstraction $(G)^{\omega}$ governing the run $r$ is the asynchronous analog of the limit inferior graph 
in the synchronous message adversary setting, as introduced in \cite{CM19:DC}. $G$ can also be viewed as  
a \emph{static} network topology, which carries all the communication occurring in some suffix of 
$r$.


%% file: conclusions.tex
\section{Conclusions}\label{sec:conclusion}

In this paper, we identified necessary and sufficient communication requirements for solving the firing rebels (FR) and firing rebels with relay (FRR) problems in asynchronous distributed systems with byzantine processes. In order to do so, we defined the concept of network abstractions, which define some of the potential message chains that must occur in a run of the system, by means of sequences of directed graphs. This way of modeling not only allows to capture communication-by-time in time-bounded asynchronous systems but also naturally translates liveness properties into connectivity properties of the graphs constituting the network abstractions. We developed novel broadcasting-based algorithms for both FR and FRR, and proved that they work correctly in all runs adhering to specific network abstractions. Moreover, we showed that it is impossible to solve FR resp.\ FRR in all runs corresponding to weaker network abstractions, which also proves that our algorithms are optimal in this respect. 

Future work will be devoted to applying our approach to other distributed computing problems.




%% file: 4-logic.tex
\section{Epistemic Logic for Byzantine Agents} \label{sec:logic}

Although presented in a purely combinatorial way, all our results have a very natural interpretation also in epistemic modal logic for byzantine distributed systems \cite{10.1007/978-3-030-29007-8_15,techreport2019,FireTark,EPTCS379.37,van2022new}. 
In the following, we will give a very brief overview of the most relevant
concepts here, but refer the reader to the general introductions 
provided in \cite{HM90} and, in particular, in the standard textbooks~\cite{FMHV03,vanDitmarsch2008}. 

We consider a set of agents $\mathcal{A}$, which represent the processes $\Pi$ in a distributed system. The configurations of the system and their relation 
are modeled as Kripke structures on the set of possible worlds $\mathcal{W}$. 
Properties of the global states, in particular, the corresponding local state
of the agents, are represented by propositional atoms. System properties, like
whether some agent has witnessed a $\START$ event, can hence be expressed via
logical formulas $\varphi$ that are to be evaluated at each of the possible worlds. 

An equivalence relation $\mathcal{R}_a$ on $\mathcal{W}$, indexed by the set of agents $\mathcal{A}$, tells which worlds are indistinguishable for an agent $a \in \mathcal{A}$. 
It facilitates the definition of the modal \emph{knowledge} operator, which tells what an agent knows in a given world:  in a Kripke model $M$ at a world $w$, agent $a$ knows $\varphi$, denoted by $M,w \vDash K_a \varphi$, if $M,w'\vDash \varphi$ for any $w'$ such that $(w,w') \in R_a$. That is, we say that $a$ knows $\varphi$ iff $\varphi$ holds at any world $w'$ that is indistinguishable from $w$ for $a$.

The runs-and-systems modeling introduced in \cite{FMHV03} adds a time evolution to
the epistemic model, by adding a global time parameter for evaluating formulas.
Our formulas hence also contain temporal modalities such as $\Diamond \varphi$ for eventually $\varphi$, and $\Box\varphi$ for always $\varphi$, in addition to epistemic modalities like $K_a$. Together, they allow us to describe and reason about the evolution of knowledge in the
runs of a distributed system.

Unfortunately, however, in system where some agents may be byzantine faulty, and where the agents can only gain knowledge through communication, the possibility of a brain-in-a-vat scenario entails that not much can be known by the agents \cite{EPTCS297.19}. Therefore, in fault-tolerant systems, knowledge must be relativized to the correctness of an agent. This yields the notion of belief as defeasible knowledge $B_i \varphi := K_a (correct_a \rightarrow \varphi)$~\cite{MOSES1993299}. 

Belief, however, does not properly capture the information gain of agent $a$ when receiving a message from $b$, which contains information encoded in some formula $\varphi$. What agent $a$ can actually infer from this message reception is $correct_b \rightarrow B_b (\varphi)$, which can be neatly expressed via the hope modality $H_a \varphi := correct_a \rightarrow B_a \varphi$ as defined by Fruzsa et.al.\ in~\cite{10.1007/978-3-030-29007-8_15}. It turns out
that message chains, i.e., the result of forwarding some information $\varphi$ via multiple hops, leads to hope chains: For instance, a message chain where process $a$ communicated $\varphi$ to $b$, and $b$ in turn relayed such a message to $c$ corresponds to $B_c (H_b (H_a (\varphi)))$. Schlögl and Schmid~\cite{EPTCS379.37} showed that $f+1$ disjoint hope chains of $\varphi$ are sufficient for acquiring belief of $\varphi$. Furthermore, hope formulas can be used to express that a process is a witness of an event like $\START$. If a propositional atom $p$ represents that $START$ occurred, then $H_a p$ can be interpreted as $a$ is a witness of $\START$.

In the following subsections, we will translate our necessary and sufficient communication abstractions for FRR to necessary and sufficient a-priori epistemic information, which can be exploited in the implementation of an FRR solution algorithm. Note carefully that \emph{a-priori} means that this information about the network network behavior is commonly known by all the agents, and already beforehand, i.e., such that it is valid for all runs of the system.

\subsection{Vertex connectivity}

Recall that the first condition in \cref{thm:firingrebelsrelay} is that there is a $2f+1$-strongly vertex connected static network abstraction that repeats infinitely often. However, we can use \cref{lem:source} to translate this condition to a liveness condition in terms of hope chains.

For convenience, we will assume that hope chains do not contain the same agent more than once. Schlögl and Schmid showed in \cite{EPTCS379.37} that, for the purposes of belief gain, one only needs to consider such hope chains. Note that the the set of all possible hope chains $\mathcal{H}$ that do not repeat agents is finite. 

Let $T = \{a_1, \ldots, a_{2f+1} \}$ be agents, with formulas $\Phi = \{\varphi_1, \ldots, \varphi_{2f+1}\} \subseteq \mathcal{F}$ to be communicated via messages, and $b \in \mathcal{A} \setminus T$ be an arbitrary agent not in $T$. 
For the purpose of FRR, the formulas in $\mathcal{F}$ are restricted to finite conjunctions of hope chains that do not repeat agents, which may only start from the atomic proposition $START$ that is true when $START$ happened at the originating agent. 
Let $F(T,\Phi,b)$ be the set of all possible sets of $2f+1$ prefix-disjoint hope chains originating at $T$ and reaching agent $b$, and $\rho$ be a map that associates each of the hope chains to the index of the source agent. Consider the following formula:

\begin{equation}
  H_{a_1} \varphi_1 \wedge \ldots \wedge H_{a_{2f+1}} \varphi_{2f+1} \rightarrow \Diamond\Bigl(\bigvee\limits_{\Sigma \in F(T,\Phi,b)} \bigwedge \limits_{\sigma \in \Sigma} H_{\sigma} (\varphi_{\rho(\sigma)})\Bigr)  
\end{equation}

This formula states that, at any given time, if $2f+1$ different agents $a_i$ hope for a formula $\varphi_i$, then eventually there will be $2f+1$ hope chains relative to each formula $\varphi_i$ from the $2f+1$ agents available at agent $b$. Since we are assuming a flooding protocol, and hope formulas are persistent, and hope chains correspond to message chains, this formula just expresses that $T$ can eventually reach $b$ through $2f+1$ prefix-disjoint paths.

Since this formula only depends on $T$, $b$, and $\Phi$, we can define a map $\Psi$ that matches each choice of $T$, $b$, and $\Phi$ to a formula $\Psi(T,\Phi,b)$. 
Thanks to \cref{lem:source}, $2f+1$ strong vertex connectivity can hence be expressed via the following formula:
\begin{equation}
    \Box( \forall T \subset \mathcal{A} . \forall \Phi \subset \mathcal{F} . \forall b \notin T. ( \vert T \vert = 2f+1 \wedge  \vert \Phi \vert = 2f+1 ) \rightarrow \Psi(T, \Phi,b)) \label{eq:strongvc}
\end{equation}

It follows from the finiteness of $\mathcal{A}$ and $\mathcal{F}$ that \cref{eq:strongvc} can be rewritten without quantifiers and cardinality checks.

\subsection{Co-roots}

The first condition in \cref{thm:firingrebelsrelay} is the existence of a $2f+1$-co-root of size $f+1$ w.r.t.\ any set $B$ of $f$ agents. Recall that this co-root is a set of $f+1$ agents not in $B$, where each agent has at least $2f+1$ in-neighbors. Let $C(B)$ be the set of all possible subsets of $\mathcal{A}$ of size $f+1$ that do not intersect $B$. Similarly, let $S(B,a)$ be the set of all possible subsets of $\mathcal{A}$ of size $2f+1$ that do not intersect $B\cup \{a\}$. Finally, let $\Phi$ be a set of $2f+1$ formulas, and $\sigma$ be an arbitrary indexing map that assigns, to any set of agents $\Delta$ of size $2f+1$, a bijection between $\Delta$ and $\Phi$. The following formula expresses that there is a co-root w.r.t $B$ of size at least $f+1$.
\begin{equation}
    \bigvee \limits_{\Sigma \in C(B)} \bigwedge \limits_{a \in \Sigma} \bigvee\limits_{\Delta \in S(B,a)} \bigwedge\limits_{b \in \Delta} H_b (\sigma(\Delta)(b)) \rightarrow \Diamond H_a H_b (\sigma(\Delta)(b)) \label{eq:coroot1}
\end{equation}

Whereas the formulas in $\mathcal{F}$ need not be hope chains, we can safely assume that the again only originate in the
atomic proposition that $START$ occurred. 

Therefore, we can reduce \cref{eq:coroot1} as follows:

\begin{equation}
    \bigvee \limits_{\Sigma \in C(B)} \bigwedge \limits_{a \in \Sigma} \bigvee\limits_{\Delta \in S(B,a)} \bigwedge\limits_{b \in \Delta} H_b (START) \rightarrow \Diamond H_a H_b (START) \label{eq:coroot2}
\end{equation}

Since \cref{eq:coroot2} depends only on $B$, we can use a mapping $G(B)$ that yields the respective formula for each possible choice of $B$, which results if the following formula:

\begin{equation}
    \Box( \forall B \subset \mathcal{A}.( \vert B \vert = f \rightarrow G(B))\label{eq:coroot3}
\end{equation}

It again follows from the finiteness of $\mathcal{A}$ that the above formula can be rewritten without quantifiers or cardinality checks.

\begin{corollary}
    Formulas \cref{eq:strongvc} and \cref{eq:coroot3} constitute the necessary and sufficient a-priori knowledge for solving FRR in a distributed system with $n$ agents, where up to $f$ may be byzantine.
\end{corollary}

%% file: main.bbl
\begin{thebibliography}{}

\end{thebibliography}


\begin{thebibliography}{10}

\bibitem{AG13}
Yehuda Afek and Eli Gafni.
\newblock Asynchrony from synchrony.
\newblock In {\em Distributed Computing and Networking}, volume 7730 of {\em Lecture Notes in Computer Science}, pages 225--239. Springer Berlin Heidelberg, 2013.

\bibitem{BCT96}
Anindya Basu, Bernadette Charron-Bost, and Sam Toueg.
\newblock Crash failures vs. crash + link failures.
\newblock In {\em Proceedings of the Fifteenth Annual ACM Symposium on Principles of Distributed Computing}, page 246, Philadelphia, Pennsylvania, United States, 1996. ACM Press.

\bibitem{BM14:JACM}
Ido Ben-Zvi and Yoram Moses.
\newblock Beyond {L}amport's happened-before: On time bounds and the ordering of events in distributed systems.
\newblock {\em J. ACM}, 61(2):13:1--13:26, April 2014.

\bibitem{CFQS12:TVG}
Arnaud Casteigts, Paola Flocchini, Walter Quattrociocchi, and Nicola Santoro.
\newblock Time-varying graphs and dynamic networks.
\newblock {\em IJPEDS}, 27(5):387--408, 2012.

\bibitem{CHARRONBOST2019100}
Bernadette Charron-Bost and Shlomo Moran.
\newblock The firing squad problem revisited.
\newblock {\em Theoretical Computer Science}, 793:100--112, 2019.

\bibitem{CM19:DC}
Bernadette Charron{-}Bost and Shlomo Moran.
\newblock Minmax algorithms for stabilizing consensus.
\newblock {\em Distributed Comput.}, 34(3):195--206, 2021.

\bibitem{10.1145/22145.22182}
B~A Coan, D~Dolev, C~Dwork, and L~Stockmeyer.
\newblock The distributed firing squad problem.
\newblock In {\em Proceedings of the Seventeenth Annual ACM Symposium on Theory of Computing}, STOC '85, page 335–345, New York, NY, USA, 1985. Association for Computing Machinery.

\bibitem{DMR08}
Yefim Dinitz, Shlomo Moran, and Sergio Rajsbaum.
\newblock Bit complexity of breaking and achieving symmetry in chains and rings.
\newblock {\em J. ACM}, 55(1):3:1--3:28, February 2008.

\bibitem{DLS88}
Cynthia Dwork, Nancy Lynch, and Larry Stockmeyer.
\newblock Consensus in the presence of partial synchrony.
\newblock {\em Journal of the ACM}, 35(2):288--323, April 1988.

\bibitem{FMHV03}
R.~Fagin, Y.~Moses, J.Y. Halpern, and M.Y. Vardi.
\newblock {\em Reasoning About Knowledge}.
\newblock MIT Press, 2003.

\bibitem{FLM86}
Michael~J. Fischer, Nancy Lynch, and Michael Merritt.
\newblock Easy impossibility proofs for distributed consensus problems.
\newblock {\em Distributed Computing}, 1(1):26--39, 1986.

\bibitem{FLP85}
Michael~J. Fischer, Nancy~A. Lynch, and M.~S. Paterson.
\newblock Impossibility of distributed consensus with one faulty process.
\newblock {\em Journal of the ACM}, 32(2):374--382, April 1985.

\bibitem{FireTark}
Krisztina Fruzsa, Roman Kuznets, and Ulrich Schmid.
\newblock Fire!
\newblock In Joseph~Y. Halpern and Andr{\'{e}}s Perea, editors, {\em Proceedings Eighteenth Conference on Theoretical Aspects of Rationality and Knowledge, {TARK} 2021, Beijing, China, June 25-27, 2021}, volume 335 of {\em {EPTCS}}, pages 139--153, 2021.

\bibitem{GWSR19:SSS}
Hugo~Rincon Galeana, Kyrill Winkler, Ulrich Schmid, and Sergio Rajsbaum.
\newblock A topological view of partitioning arguments: Reducing k-set agreement to consensus.
\newblock In {\em Stabilization, Safety, and Security of Distributed Systems - 21st International Symposium, {SSS} 2019, Pisa, Italy, October 22-25, 2019, Proceedings}, volume 11914 of {\em Lecture Notes in Computer Science}, pages 307--322. Springer, 2019.

\bibitem{GP16:OPODIS}
Emmanuel Godard and Eloi Perdereau.
\newblock k-set agreement in communication networks with omission faults.
\newblock In Panagiota Fatourou, Ernesto Jim{\'{e}}nez, and Fernando Pedone, editors, {\em 20th International Conference on Principles of Distributed Systems, {OPODIS} 2016, December 13-16, 2016, Madrid, Spain}, volume~70 of {\em LIPIcs}, pages 8:1--8:17. Schloss Dagstuhl - Leibniz-Zentrum f{\"{u}}r Informatik, 2016.

\bibitem{GM20:JACM}
Guy Goren and Yoram Moses.
\newblock Silence.
\newblock {\em J. ACM}, 67(1), jan 2020.

\bibitem{HM90}
Joseph~Y. Halpern and Yoram Moses.
\newblock Knowledge and common knowledge in a distributed environment.
\newblock {\em J. ACM}, 37(3):549--587, 1990.

\bibitem{vanDitmarsch2008}
Barteld~Kooi Hans~van Ditmarsch, Wiebe~Hoek.
\newblock {\em Dynamic Epistemic Logic}, pages 11--42.
\newblock Springer Netherlands, Dordrecht, 2008.

\bibitem{HW12:PODC}
Stephan Holzer and Roger Wattenhofer.
\newblock Optimal distributed all pairs shortest paths and applications.
\newblock In {\em Proceedings of the 2012 ACM Symposium on Principles of Distributed Computing}, PODC '12, pages 355--364, New York, NY, USA, 2012. ACM.

\bibitem{KO11:SIGACT}
F.~Kuhn and R.~Oshman.
\newblock Dynamic networks: Models and algorithms.
\newblock {\em SIGACT News}, 42(1):82--96, 2011.

\bibitem{KN97}
Eyal Kushilevitz and N.~Nisan.
\newblock {\em Communication Complexity}.
\newblock Cambridge University Press, 1997.

\bibitem{EPTCS297.19}
Roman Kuznets, Laurent Prosperi, Ulrich Schmid, and Krisztina Fruzsa.
\newblock Causality and epistemic reasoning in byzantine multi-agent systems.
\newblock In Lawrence~S. Moss, editor, {\em {\rm Proceedings Seventeenth Conference on} Theoretical Aspects of Rationality and Knowledge, {\rm Toulouse, France, 17-19 July 2019}}, volume 297 of {\em Electronic Proceedings in Theoretical Computer Science}, pages 293--312. Open Publishing Association, 2019.

\bibitem{10.1007/978-3-030-29007-8_15}
Roman Kuznets, Laurent Prosperi, Ulrich Schmid, and Krisztina Fruzsa.
\newblock Epistemic reasoning with byzantine-faulty agents.
\newblock In Andreas Herzig and Andrei Popescu, editors, {\em Frontiers of Combining Systems}, pages 259--276, Cham, 2019. Springer International Publishing.

\bibitem{techreport2019}
Roman Kuznets, Laurent Prosperi, Ulrich Schmid, Krisztina Fruzsa, and Lucas Gréaux.
\newblock Knowledge in byzantine message-passing systems i framework and the causal cone.
\newblock Eprint, TU Wien, 2019.

\bibitem{Lam78}
Leslie Lamport.
\newblock Time, clocks, and the ordering of events in a distributed system.
\newblock {\em Commun. ACM}, 21(7):558--565, 1978.

\bibitem{LSP82}
Leslie Lamport, Robert Shostak, and Marshall Pease.
\newblock The byzantine generals problem.
\newblock {\em ACM Trans. Program. Lang. Syst.}, 4(3):382–401, jul 1982.

\bibitem{10.1145/2591796.2591853}
Hammurabi Mendes, Christine Tasson, and Maurice Herlihy.
\newblock Distributed computability in byzantine asynchronous systems.
\newblock In {\em Proceedings of the Forty-Sixth Annual ACM Symposium on Theory of Computing}, STOC '14, page 704–713, New York, NY, USA, 2014. Association for Computing Machinery.

\bibitem{MOSES1993299}
Yoram Moses and Yoav Shoham.
\newblock Belief as defeasible knowledge.
\newblock {\em Artificial Intelligence}, 64(2):299--321, 1993.

\bibitem{10.1145/3293611.3331624}
Thomas Nowak, Ulrich Schmid, and Kyrill Winkler.
\newblock Topological characterization of consensus under general message adversaries.
\newblock In {\em Proceedings of the 2019 ACM Symposium on Principles of Distributed Computing}, PODC '19, page 218–227, New York, NY, USA, 2019. Association for Computing Machinery.

\bibitem{PS18:SIROCCO}
Daniel Pfleger and Ulrich Schmid.
\newblock On knowledge and communication complexity in distributed systems.
\newblock In {\em Structural Information and Communication Complexity - 25th Intern ational Colloquium, {SIROCCO} 2018, Ma'ale HaHamisha, Israel, June 18-21, 2018, Revised Selected Papers}, pages 312--330, 2018.

\bibitem{RS11:TCS}
Peter Robinson and Ulrich Schmid.
\newblock The {A}synchronous {B}ounded-{C}ycle model.
\newblock {\em Theoretical Computer Science}, 412(40):5580--5601, 2011.

\bibitem{EPTCS379.37}
Thomas Schlögl and Ulrich Schmid.
\newblock A sufficient condition for gaining belief in byzantine fault-tolerant distributed systems.
\newblock In Rineke Verbrugge, editor, {\em {\rm Proceedings Nineteenth conference on} Theoretical Aspects of Rationality and Knowledge, {\rm Oxford, United Kingdom, 28-30th June 2023}}, volume 379 of {\em Electronic Proceedings in Theoretical Computer Science}, pages 487--506. Open Publishing Association, 2023.

\bibitem{SSW18:SIROCCO}
Ulrich Schmid, Manfred Schwarz, and Kyrill Winkler.
\newblock On the strongest message adversary for consensus in directed dynamic networks.
\newblock In {\em Structural Information and Communication Complexity - 25th International Colloquium, {SIROCCO} 2018, Ma'ale HaHamisha, Israel, June 18-21, 2018, Revised Selected Papers}, pages 102--120, 2018.

\bibitem{SS21:SSS}
Manfred Schwarz and Ulrich Schmid.
\newblock Round-oblivious stabilizing consensus in dynamic networks.
\newblock In Colette Johnen, Elad~Michael Schiller, and Stefan Schmid, editors, {\em Stabilization, Safety, and Security of Distributed Systems - 23rd International Symposium, {SSS} 2021, Virtual Event, November 17-20, 2021, Proceedings}, volume 13046 of {\em Lecture Notes in Computer Science}, pages 154--172. Springer, 2021.

\bibitem{ST87}
T.~K. Srikanth and Sam Toueg.
\newblock Simulating authenticated broadcasts to derive simple fault-tolerant algorithms.
\newblock {\em Distributed Computing}, 2(2):80–94, June 1987.

\bibitem{TV15:PODC}
Lewis Tseng and Nitin~H. Vaidya.
\newblock Fault-tolerant consensus in directed graphs.
\newblock In {\em Proceedings of the 2015 ACM Symposium on Principles of Distributed Computing}, PODC '15, page 451–460, New York, NY, USA, 2015. Association for Computing Machinery.

\bibitem{van2022new}
Hans van Ditmarsch, Krisztina Fruzsa, and Roman Kuznets.
\newblock A new hope.
\newblock {\em Advances in Modal Logic}, 14:349--369, 2022.

\bibitem{IntroGraphTh}
Douglas~B. West.
\newblock {\em Introduction to Graph Theory}.
\newblock Prentice Hall, 2001.

\bibitem{WS07:DC}
Josef Widder and Ulrich Schmid.
\newblock Booting clock synchronization in partially synchronous systems with hybrid process and link failures.
\newblock {\em Distributed Computing}, 20(2):115--140, August 2007.

\bibitem{WS09:DC}
Josef Widder and Ulrich Schmid.
\newblock The {T}heta-{M}odel: Achieving synchrony without clocks.
\newblock {\em Distributed Computing}, 22(1):29--47, April 2009.

\bibitem{winkler_et_al:LIPIcs.ITCS.2023.100}
Kyrill Winkler, Ami Paz, Hugo Rincon~Galeana, Stefan Schmid, and Ulrich Schmid.
\newblock {The Time Complexity of Consensus Under Oblivious Message Adversaries}.
\newblock In Yael Tauman~Kalai, editor, {\em 14th Innovations in Theoretical Computer Science Conference (ITCS 2023)}, volume 251 of {\em Leibniz International Proceedings in Informatics (LIPIcs)}, pages 100:1--100:28, Dagstuhl, Germany, 2023. Schloss Dagstuhl -- Leibniz-Zentrum f{\"u}r Informatik.

\bibitem{WSM19:OPODIS}
Kyrill Winkler, Ulrich Schmid, and Yoram Moses.
\newblock A characterization of consensus solvability for closed message adversaries.
\newblock In {\em 23rd International Conference on Principles of Distributed Systems, {OPODIS}}, volume 153 of {\em LIPIcs}, pages 17:1--17:16. Schloss Dagstuhl - Leibniz-Zentrum f{\"{u}}r Informatik, 2019.

\bibitem{WSS19:DC}
Kyrill Winkler, Manfred Schwarz, and Ulrich Schmid.
\newblock Consensus in directed dynamic networks with short-lived stability.
\newblock {\em Distributed Computing}, 32(5):443--458, 2019.

\bibitem{Yao79}
Andrew~Chi{-}Chih Yao.
\newblock Some complexity questions related to distributive computing (preliminary report).
\newblock In {\em Proceedings of the 11h Annual {ACM} Symposium on Theory of Computing, April 30 - May 2, 1979, Atlanta, Georgia, {USA}}, pages 209--213, 1979.

\bibitem{COULOUMA201580}
Étienne Coulouma, Emmanuel Godard, and Joseph Peters.
\newblock A characterization of oblivious message adversaries for which consensus is solvable.
\newblock {\em Theoretical Computer Science}, 584:80--90, 2015.
\newblock Special Issue on Structural Information and Communication Complexity.

\end{thebibliography}
